\pdfoutput=1
\documentclass[mnsc,nonblindrev]{informs3} % current default for manuscript submission

\OneAndAHalfSpacedXI
\usepackage{amsmath}
\usepackage{bm}
\usepackage{caption}
\usepackage{threeparttable}
\usepackage{float}
\usepackage{booktabs}  
\usepackage{chngcntr}
\usepackage{xcolor}
\usepackage{tabularx}
\usepackage{multirow}
\usepackage{dcolumn}
\usepackage{pifont}
\newcolumntype{.}{D{.}{.}{-1}}
\usepackage{adjustbox}
\usepackage{siunitx}
\usepackage{lscape}
\usepackage{etoolbox}
\usepackage{graphics}
\usepackage{subfig}
\usepackage{appendix}
\usepackage{verbatim}
\appto\TPTnoteSettings{\footnotesize}
\usepackage{natbib}
 \bibpunct[, ]{(}{)}{,}{a}{}{,}%
 \def\bibfont{\small}%
\usepackage[colorlinks=true,
linkcolor=blue,
urlcolor=blue,
citecolor=blue]{hyperref}

\TheoremsNumberedThrough     % Preferred (Theorem 1, Lemma 1, Theorem 2)
\ECRepeatTheorems

\EquationsNumberedThrough    % Default: (1), (2), ...
\MANUSCRIPTNO{MS-INS-20-03320}

\begin{document}
%%%%%%%%%%%%%%%%

%\setcounter{page}{1}
%\DOI{10.1287/xxxx.0000.0000}%


\RUNTITLE{Emotions in Online Content Diffusion}

% \TITLE{Bundling Information Goods of Decreasing Value}
\TITLE{Emotions in Online Content Diffusion}

\ARTICLEAUTHORS{%
\AUTHOR{Yifan Yu\thanks{These authors contributed equally to this work.}}
\AFF{Michael G. Foster School of Business, University of Washington, Seattle, Washington 98195}

\AUTHOR{Shan Huang\footnotemark[1]\thanks{To whom correspondence should be addressed. E-mail: shanhh@hku.hk}}
\AFF{Faculty of Business and Economics, The University of Hong Kong, Hong Kong}

\AUTHOR{Yuchen Liu and Yong Tan}
\AFF{Michael G. Foster School of Business, University of Washington, Seattle, Washington 98195}
%, \URL{}}
% \AUTHOR{Marg Arinella}
% \AFF{Institute for Food Adulteration, University of Food Plains, Food Plains, MN 55599, \EMAIL{m.arinella@adult.ufp.edu}}
} % end of the block

\ABSTRACT{%
Social media-transmitted online information, which is associated with emotional expressions, shapes our thoughts and actions. In this study, we incorporate social network theories and analyses and use a computational approach to investigate how emotional expressions, particularly \textit{negative discrete emotional expressions} (i.e., anxiety, sadness, anger, and disgust), lead to differential diffusion of online content in social media networks. We rigorously quantify diffusion cascades' structural properties (i.e., size, depth, maximum breadth, and structural virality) and analyze the individual characteristics (i.e., age, gender, and network degree) and social ties (i.e., strong and weak) involved in the cascading process. In our sample, more than six million unique individuals transmitted 387,486 randomly selected articles in a massive-scale online social network, WeChat. We detect the expression of discrete emotions embedded in these articles, using a newly generated domain-specific and up-to-date emotion lexicon. We apply a partial-linear instrumental variable approach with a double machine learning framework to causally identify the impact of the negative discrete emotions on online content diffusion. We find that articles with more expressions of anxiety spread to a larger number of individuals and diffuse more deeply, broadly, and virally. Expressions of anger and sadness, however, reduce cascades' size and maximum breadth. We further show that the articles with different degrees of negative emotional expressions tend to spread differently based on individual characteristics and social ties. Our results shed light on content marketing and regulation, utilizing negative emotional expressions.

% We are among the first to investigate how discrete negative emotional expressions affect information spread through different individuals and social ties, leading to differential diffusion outcomes in a massive-scale social media network.

% Our results show that readers well received the emotions expressed in the articles, since consistent emotions were detected between articles and the associated comments. Our paper has rich implications for promoting information (e.g., articles, products and elections) in social networks.
% Enter your abstract
% We detected the degree of eight discrete emotions (i.e., surprise, joy, anticipation, love, anxiety, sadness, anger, and disgust) embedded in the content of each article with a newly generated domain-specific and up-to-date emotion lexicon. 
}%

%\KEYWORDS{deterministic inventory theory; infinite linear programming duality;
%  existence of optimal policies; semi-Markov decision process; cyclic schedule}

%\HISTORY{This paper was first submitted on April 12, 1922 and has been with the authors for 83 years for 65 revisions.}

\maketitle
%%%%%%%%%%%%%%%%%%%%%%%%%%%%%%%%%%%%%%%%%%%%%%%%%%%%%%%%%%%%%%%%%%%%%%

% Samples of sectioning (and labeling) in MNSC
% NOTE: (1) \section and \subsection do NOT end with a period
%       (2) \subsubsection and lower need end punctuation
%       (3) capitalization is as shown (title style).
%
%\section{Introduction.}\label{intro} %%1.
%\subsection{Duality and the Classical EOQ Problem.}\label{class-EOQ} %% 1.1.
%\subsection{Outline.}\label{outline1} %% 1.2.
%\subsubsection{Cyclic Schedules for the General Deterministic SMDP.}
%  \label{cyclic-schedules} %% 1.2.1
%\section{Problem Description.}\label{problemdescription} %% 2.

% Text of your paper here

\section{Introduction}
Emotions are commonly expressed in natural language and shape our daily communications. Neuroscientists have found that humans often pay special attention to and establish enhanced memory for emotional events, which activate subsequent actions \citep{dolan2002emotion}. Social media, including Twitter, Facebook, and WeChat, has facilitated rapid information sharing and large-scale information cascades. Online information transmitted in social media networks, particularly content that is emotionally charged, shapes our thoughts and actions. Specifically, the emotionally charged content that spreads on social media affects our notions of such concerns as morality, ideology \citep{brady2017emotion}, politics, terrorism \citep{vosoughi2018spread},
%\citep{clarke2006emotion,vosoughi2018spread}, 
and financial investments \citep{bollen2011twitter,nguyen2020influence}. Notably, when misinformation is incorporated with the emotional expression that facilitates network diffusion, it can be harmful, especially in the case of COVID-19. Thus, it is timely and vital to understand how emotional expression affects online content diffusion in social media networks. 
% In this paper, negative emotional expressions embedded in online content are interchangeably referred to as negative emotions in content.

% These days, various news about the Covid outbreak and the vaccines, which is causing increased stress and anxiety, has filled social media and spread rapidly and widely in social networks. 

% Although considerable attention has been paid to emotions in social transmission of information \citep{berger2012makes,brady2017emotion}, existing research on the role of emotions in information diffusion often takes a dyadic perspective, and relies largely on small samples \citep{berger2012makes}, lab experiments \citep{berger2012makes, zhang2018emotional}, and short messages \citep{brady2017emotion}. 

%1. focus on the negative emotions; 2. there are 8 discrete emotions in human XXXX. In our analysis....
It is empirically challenging to rigorously characterize the process of peer-to-peer spreading of online content diffused in social media networks and, thus, to investigate the relationship between emotional expression and the structural as well as demographic properties and social ties of diffusion cascades. Understanding the characteristics of the diffusion process is critical in informing the strategies for content providers to maximize online content spreading, such as by seeding in different individuals and structural elements of networks and utilizing different models, i.e., viral or broadcast \citep{goel2015structural}. It also helps social media platforms to design policies that regulate the diffusion of unwanted information, such as negative emotional content. 
Further, the affective dimensions of content, such as embedded emotional expressions, have received much less attention than have their cognitive counterparts \citep{vosoughi2018spread}
%\citep{vosoughi2018spread, aral2016unpacking} 
due mainly to the difficulty of detecting emotional expressions, particularly in a massive amount of content. Prior studies focus mainly on the impact of emotions on the general popularity of news or tweets and take a dyadic perspective \citep[e.g.,][]{berger2012makes,brady2017emotion}. To the best of our knowledge, few comprehensive empirical investigations have examined the relationship between embedded emotional expressions and the structural, demographic and social relational properties of online content cascades in social media networks. 

We therefore conduct a very large-scale field analysis to understand whether and how emotional expressions, particularly \textit{negative emotional expressions}, embedded in online content, lead to differential diffusion patterns in a massive-scale online social network, using a computational approach. We incorporate social network theories and analysis and focus on a random sample of 387,486 online articles and their diffusion cascades in China’s largest social networking site, WeChat. We rigorously quantify and characterize diffusion cascades' structural (i.e., size, depth, maximum breadth, and structural virality) properties. We further analyze the characteristics of individuals (i.e., age, gender, and network degree) and social ties (i.e., strong and weak ties) involved in the cascading process in which more than six million unique individuals transmitted these articles in WeChat's social network.
We focus on four negative discrete emotions: anxiety, sadness, anger, and disgust, and adopt \textit{discrete emotions}\footnote{We use eight discrete emotions: surprise, joy, anticipation, love, anxiety, sadness, anger, and disgust, to delineate complex emotions. As \cite{quan2010blog} found, these eight discrete emotions are the most commonly seen in articles in China and cover the most common emotions in our context. These emotions also are considered to be the most basic emotions in discrete emotion research \citep{tomkins1962affect,plutchik1980emotion,lerner2004heart,lerner2015emotion,yin2014anxious}.} to delineate complex emotions.
According to discrete emotion theory \citep{tomkins1962affect}, there is a small number of core emotions that constitute other emotions. For example, awe may be viewed as a blend of anxiety and love. Once we understand discrete emotions, we can understand more complex emotions \citep{tomkins1962affect, plutchik1980emotion, lerner2004heart,lerner2015emotion}. Empirical evidence also shows that discrete emotions are relatively independent of each other and can better delineate human emotions than can valence (i.e., positive and negative polarity) or arousal (also “activation” or “intensity,” the extent to which a person is energized by an experience) \citep{lerner2004heart,yin2017keep,yu2019emotions}. 
To accurately detect the expressions of various discrete emotions, embedded in the content of our sampled articles, we construct a new domain-specific and up-to-date emotion lexicon, using a state-of-the-art lexicon-generation approach \citep{xue2014study,yu2019emotions}, and validate it with human annotation. 
%%We find that the correlations between the expressions of any two discrete emotions embedded in each article in our sample are generally below 0.20, confirming the independence of these emotions.

The large spread of online content with negative emotional expressions can have a profound impact. Negativity bias theory argues that negative emotional content can have a more significant impact on individuals' cognition and content consumption behavior than can positive content \citep{stieglitz2013emotions,hennig2015does}. 
% Recent literature has indicated the significant potential of negative emotional expressions in driving information diffusion, for example, in the context of moralized content \citep{brady2017emotion,brady2019ideological} and cancer tweets \citep{wang2020negative}. 
Further, as the content with negative emotional expressions permeates social media and significantly affects the opinions and mood of the public \citep{heath2001emotional}, leading to problematic consequences, it is vital to regulate the large spread of content with negative emotional expressions.\footnote{For example, it is well-documented that the media may provoke irrational anger and fear to skew laws and public policies toward trivial but emotional issues and away from legitimate but less emotional concerns \citep{bailis1996estimating,edelman1992professional}.} Understanding the impact and the mechanisms of negative emotional expressions on online content diffusion will provide critical insight into designing appropriate policies. Social psychology theories and empirical work in management research have consistently found that people tend to share content with positive emotional expressions, because positive content reflects positively on the sender, and recipients enjoy consuming such content (e.g., see literature review by \cite{heath2001emotional} and \cite{berger2012makes}).  %\citep{taylor1988illusion,weinstein1980unrealistic,heath2001emotional,berger2012makes}. 
The impact of negative discrete emotional expressions on content diffusion, however, is still not clearly understood. Researchers have attempted to understand the effect of negative emotional expressions from the perspective of arousal, but laboratory studies have yielded contradictory results. In particular, some lab studies have shown that high-arousal negative emotional expressions (anger and anxiety) cause online news articles to go viral \citep{berger2012makes}. Other research shows that the expression of a low-arousal negative emotion (disgust) helps content such as an urban legend to diffuse in social networks, whereas high-arousal emotional expressions (anger and fear) negatively affect diffusion \citep{heath2001emotional}.

% negative emotions as negative emotional expressions embedded in online content. 
We apply a partial-linear instrumental variable approach, using a double machine learning framework \citep{chernozhukov2018double}, to identify the impact of negative emotional expressions on online content diffusion. This framework enables us not only to comprehensively control the possibly non-linear effects of the observable publisher- and article-level characteristics through incorporating machine learning techniques, but also to effectively address the endogeneity concern of emotional expressions induced by unobservable factors. 
We find that expressions of anxiety exhibit the most positive impact on online content diffusion, whereas expressions of anger and sadness adversely affect diffusion. More specifically, expressions of anxiety embedded in content significantly increase all of the cascade dimensions, including depth, size, breadth, and structural virality. In contrast, expressions of anger and sadness significantly decrease cascade size and breadth, and have an insignificant impact on cascade depth and structural virality. Expressions of disgust, however, do not show significant effects on cascade depth, size, breadth, or structural virality.

We further analyze the demographic characteristics (i.e., age, gender, and network degree) of more than six million individuals involved in the cascading process and the social ties (i.e., weak and strong) among them. We find that all four types of negative emotional expressions play significant roles in affecting the cascading process, based on the demographic and network characteristics of users who participate in the cascades. First, cascades of articles with more expressions of anger and anxiety are shared by older users, whereas younger users are engaged in sharing articles with more expressions of disgust. Second, articles with more expressions of anxiety and disgust are spread by users with a greater network degree (with more friends in their local social networks), whereas articles with more expressions of anger are spread by users with a smaller network degree (with fewer friends). Third, articles with more expressions of anger and anxiety are spread more through strong ties. Theories and empirical evidence suggest that weak ties and individuals of a larger network degree facilitate network diffusion \citep{granovetter1977strength,banerjee2013diffusion}. This understanding, together with our findings, may explain the negative impact of expressions of anger, which tend to diffuse among those with strong ties and users with fewer friends, inhibiting the content spread. Finally, females participate less than do males in the cascades of articles that have more expressions of sadness.

Our work makes several novel contributions. First, it adds to the emerging literature on peer-to-peer spread of online content in social media networks and its economic and social implications \citep[e.g.,][]{goel2015structural,vosoughi2018spread}. Our study is timely and speaks to this expanding area of inquiry. 
In particular, our research provides some of the first large-scale comprehensive empirical evidence on how negative emotional expressions shape the diffusion of online information. 
Second, we accurately measure the structural properties of the cascades of a large number of online articles. Our work goes beyond a dyadic perspective of content sharing \citep{berger2012makes} and demonstrates that the expressions of various discrete negative emotions lead to differential cascading structures and processes. 
Third, we are among the first to uncover the relationship between emotional expressions in content and the individual characteristics of a cascade's participants. We thus not only provide novel empirical evidence and insight but also inform the strategies to promote or inhibit content diffusion in online social networks. 
Fourth, leveraging a semi-parametric instrumental variable approach, we make causal inferences based on large-scale observational data, differing from past studies built on predictive models \citep{berger2012makes,stieglitz2013emotions,brady2017emotion,wang2020negative} or small-scale lab experiments \citep{heath2001emotional,berger2012makes}. 
Finally, the existing theories of emotions and online content diffusion \citep{berger2012makes,stieglitz2013emotions,hennig2015does} cannot sufficiently theorize the effects of different negative emotional expressions in our results. Therefore, we further contribute to the literature by proposing a customized EASI framework \citep{van2009emotions}, which incorporates the theories of emotional contagion \citep[e.g.,][]{hatfield1993emotional,barsade2002ripple,del2016echo}, emotion-induced approach-avoidance motivation \citep{elliot2013approach}, cognitive appraisal \citep{lerner2015emotion}, and social comparison \citep{heath2001emotional}, some of the most important theories of emotional expressions in the social psychology literature.

\section{Literature}
% Our work is grounded in theories of emotion, social networks, and information diffusion, and relates to the emerging research on the spread of online content and its social and economic implications \citep[e.g.,][]{stieglitz2013emotions,shi2014content,goel2015structural, brady2017emotion,vosoughi2018spread}. In Section \ref{lr:de}, we review the literature on emotion theories and their application in management research. In Section \ref{lr:negative}, we review the literature on emotions in online content diffusion. 
% In Section \ref{lr:cv}, we discuss the impact of negative emotional expressions on online content diffusion based on theories and findings in previous literature.

\subsection{Discrete Emotions} \label{lr:de}
% Understanding the social and economic implications of emotions embedded in online content has been an important topic in current management research \citep{zhang2013affective,yin2014anxious}. This stream of literature

Management research that examines the impact of emotional expressions embedded in online content on user behavior \citep[e.g.,][]{hennig2015does,yin2017keep,song2019using} is driven primarily by dimensional emotion theory \citep{russell1980circumplex}, which explains the effect of emotions using dimensions such as valence and arousal. Prior research shows that valence in online reviews affects individuals’ product evaluation and adoption \citep{hennig2015does}, and aggregated product sales \citep{song2019using},
%Prior research shows that valence in online reviews affects individuals’ product evaluations \citep{doh2009consumers}, intention to recommend products \citep{lee2009electronic}, perceived helpfulness of reviews \citep{hong2017understanding}, product early adoptions \citep{hennig2015does}, and daily sales \citep{song2019using}, 
whereas arousal is associated with review helpfulness \citep{yin2017keep}. \cite{stieglitz2013emotions} find that positive and negative (valence) tweets tend to be retweeted more often and more quickly than is neutral content. \cite{berger2012makes} find that news with more positive and high-arousal emotional expressions is more likely to be shared by readers. Recent psychological research, however, finds that emotions with similar valence and arousal can lead to divergent effects on people’s decision making and judgment \citep{lerner2015emotion}. For example, 
%disgust and sadness are both negative emotions. Disgust, however, reduces a consumer’s willingness to pay, whereas sadness increases it \citep{lerner2004heart}. 
anger and anxiety are similar in both valence and arousal, but anxiety embedded in online reviews induces higher perceived helpfulness than does anger \citep{yin2014anxious}. These findings suggest that the dimensions of valence and arousal cannot fully explain the variations in the effects of emotions \citep{plutchik2001nature,lerner2015emotion}. 

Recent work uses discrete emotion theory to measure emotional expressions in online content \citep[e.g.,][]{malik2017helpfulness,yu2019emotions,nguyen2020influence}. Discrete emotion theory, a competing theory for the dimensional emotion model, identifies specific basic and independent emotions, i.e., discrete emotions, that constitute other human emotions \citep{tomkins1962affect,plutchik1980emotion}. Discrete emotions are rooted in human evolution, with expression and recognition fundamentally as the same across all individuals, regardless of ethnic or cultural differences \citep{plutchik2001nature}. The theory originated from the late 19th-century evolutionary theory of Charles Darwin, who argued that certain basic emotions evolved from natural selection. Neuroimaging analyses have shown that these discrete emotions are linked to discrete neural signatures and certain structures of the human brain \citep{saarimaki2016discrete}. Prior management research shows that discrete emotional expressions in online word-of-mouth are of larger predictive power than is valence for perceived review helpfulness \citep{yin2014anxious,malik2017helpfulness}, consumers' purchase decisions and sales performance \citep{yu2019emotions}, and stock market returns \citep{nguyen2020influence}. Thus, discrete emotions provide a means to understand the effects of emotional expressions in online content diffusion. We focus on four negative discrete emotions (i.e., anger, anxiety, sadness, and disgust) and control for four positive discrete emotions (i.e., love, joy, surprise, and anticipation). The eight discrete emotions used in our work are the most commonly expressed ones in online content \citep{quan2010blog,yu2019emotions}. These emotions also are considered to be the most basic emotions in discrete emotion research \citep{tomkins1962affect,plutchik1980emotion,lerner2004heart,lerner2015emotion,yin2014anxious}. 

\subsection{Emotions in Online Content Diffusion} \label{lr:negative}
%Understanding what drives online content diffusion has attracted the increasing efforts of researchers from various fields \citep[e.g.,][]{banerjee2013diffusion,ransbotham2013impact,shi2014content,mitra2015information,vosoughi2018spread}. Notably, 
Emotional expressions embedded in content are critical in the social transmission of online information \citep{brady2017emotion,brady2019ideological}. Individuals are motivated to share emotionally charged content to strengthen social connections, coordinate actions, build a persona, reduce emotion-related ambiguous sensations, and rationalize their emotional experiences \citep{berger2012makes}.
%Individuals are motivated to share emotionally charged content to strengthen social connections, coordinate actions \citep{peters2007social}, build a persona \citep{berger2012makes}, reduce emotion-related ambiguous sensations and rationalize their emotional experiences \citep{rime1991beyond}. 
Our knowledge of how emotional expressions shape general online content diffusion in large-scale social media networks, however, is still limited. Prior studies mostly focus on a specific type of online content, such as \textit{New York Times} articles shared by email \citep{berger2012makes} and political \citep{stieglitz2013emotions}, moralized \citep{brady2017emotion,brady2019ideological}, or cancer-related \citep{wang2020negative} tweets. Instead, our sample is comprehensive and representative (randomly collected), covers a variety of topics, and includes both short texts (similar to tweets) and long articles (similar to newspaper and magazine articles). 

The detection of emotional expressions in a very large amount of content is a recent development of Natural Language Processing (NLP). Early research that relies on small-scale lab surveys to measure disgust \citep{heath2001emotional} and arousal \citep{berger2012makes} has been criticized for containing self-report biases and is subject to measurement errors. 
More recent studies use predefined and domain-independent emotion lexicons to detect emotion words in content \citep{stieglitz2013emotions,brady2017emotion,wang2020negative}. These lexicons do not account for the contextual differences of emotional expressions and capture only a fraction of total emotion words in online content \citep{xue2014study,yin2014anxious,yu2019emotions}. 
Our study advances the measurement of emotional expressions in online content, through its implementation of a scalable and domain-adaptive computational approach to detect emotional expressions embedded in content \citep{song2018directional,yu2019emotions}. Notably, our analysis shows that predefined and domain-independent emotion lexicons miss 58.4\% of the unique emotion words in our sampled articles (see Appendix \ref{appendix:construct}). 

The general popularity of content, such as a dummy indicator of whether an article makes the \textit{New York Times}’ most e-mailed list \citep{berger2012makes} or the number of retweets \citep{stieglitz2013emotions,brady2017emotion,brady2019ideological,wang2020negative}, is the most used measure of online content diffusion in the literature. These measures, however, are not sufficient to describe information diffusion in large-scale social media networks. We therefore use multiple aspects of the diffusion cascades: structural properties, demographics, and social ties, and collect a large-scale comprehensive sample of online content and their diffusion cascades from a major social media network. 
We measure the structural properties of diffusion cascades through four dimensions: size (the total number of users involved in sharing the article), depth (the maximum number of sharing hops from the original article, for which a hop is a sharing action by a new unique user), maximum breadth (the maximum number of users involved in the cascade at any depth), and structural virality (the average length of the shortest paths between all pairs of nodes in a diffusion tree) \citep{goel2015structural,vosoughi2018spread}. We further investigate how emotional expressions affect the demographics and social ties of cascades, which is critical to designing seeding strategies to maximize the diffusion of online content in social networks. To the best of our knowledge, few studies in management literature have explored these relationships empirically, or proposed integrated frameworks to theorize the relationships. 
Only indirect predictions can be made based on the social psychology literature (as detailed in Section \ref{lr:cv}).

Different structural dimensions of a cascade can provide an understanding and critical implications of information diffusion in social networks. Size is the general popularity of an article, without specifying its diffusion structure. Depth and maximum breadth are standard summary statistics of the structure of a cascade. Intuitively, depth is the generations of a cascade, indicating the maximum degree of contacts that the information can reach from the root node. Information with a cascade of a higher level of depth is more likely to break into different social communities. Maximum breadth provides an intuition about how ``broad'' or ``wide'' the cascade is, which often results from broadcast structures. A ``broad'' but not ``deep'' cascade implies that the content is spread within the same but a large social community. Structure virality quantifies the distinction between single broadcast and viral diffusion \citep{goel2015structural}. Higher structure virality indicates that the cascade is driven more by decentralized and peer-to-peer sharing than by broadcasting. Broadcasting and viral diffusion are two typical alternative ways that enable information to reach large audiences \citep{van2018customer}. The cascade structures can be influenced by diffusion strategies, such as whether to utilize mass media and high-centrality users to broadcast content or use rewards to encourage peer-to-peer diffusion. Moreover, how to promote an article or product in networks also correlates with potential cascade structures. 

Finally, in the existing literature, the relationship between emotional expressions and online content diffusion is built mainly on observational studies, using predictive models \citep{berger2012makes,stieglitz2013emotions,brady2017emotion,wang2020negative}, and small-scale lab experiments \citep{heath2001emotional,berger2012makes}. Although prior observational studies control for the observed confounding factors, they may not identify the causation; rather, mainly the correlation between emotional expressions and online content diffusion. For example, \cite{berger2012makes} examines the correlation between emotional expressions and whether a news article went viral through emails and applies lab experiments to back up the causalities between emotions and information sharing. \cite{brady2017emotion} uses a multilevel regression model to investigate the association between moral-emotional words in tweets and the number of retweets. \cite{stieglitz2013emotions} and \cite{wang2020negative} adopt the ordinary least squares regression model to predict the total number of retweets with emotional expressions. Without causal identifications,  omitted variables (e.g., writers' fixed effects, unobserved content characteristics) and measurement errors in emotional content may both bias the estimation. It is also possible that the impact of control variables is non-linear, violating the linearity assumptions of the models. 
Further, lab studies have been criticized due to their self-report biases and lack of external validity. By using a semi-parametric instrumental variable approach that incorporates machine learning techniques \citep{chen2016xgboost,chernozhukov2018double}, our study is among the first to causally identify the impact of emotional expressions on online content diffusion with a large-scale and real-world sample of online content.

\section{Theory: Impact of Negative Emotional Expressions on Online Content Diffusion}
%The Impact of Negative Emotions on Online Content Diffusion
\label{lr:cv}
% The existing theoretical frameworks, i.e., social comparison \citep{heath2001emotional}, emotional contagion \citep{brady2017emotion,brady2019ideological,wang2020negative}, and negativity bias \citep{stieglitz2013emotions,wang2020negative}, have their limitations in explaining the heterogeneous effects of negative discrete emotional expressions on online content diffusion. The social comparison theory only applies for the effect of disgust \citep{heath2001emotional}. The emotional contagion theory explains the diffusion of emotional states \citep{hatfield1993emotional,barsade2002ripple}, but does not focus on how emotional expressions shape online content diffusion (more discussions in Section \ref{lr:cv}). The negativity bias theory highlights the important roles of negative emotional expressions in content diffusion \citep{stieglitz2013emotions}, but does not help explain the heterogeneous effects of negative discrete emotions. Our work customizes the emotion as social information framework \citep{van2009emotions,van2010emerging}, a generalized framework that incorporates but not limited to the theories of social comparison and emotional contagion (detailed in Section \ref{lr:cv}), to theorize how and why each type of negative emotional expressions shape online content diffusion, respectively.
Although negative bias theory predicts that tweets with negative valence tend to be retweeted more readily than are neutral tweets \citep{stieglitz2013emotions}, there is no theoretical framework that can well explain the heterogeneous impact of different negative discrete emotions on online content diffusion. In the process of cognition and decision making, different negative discrete emotions show distinct effects \citep[e.g.,][]{lerner2004heart,yin2014anxious,lerner2015emotion,yu2019emotions}. Therefore, it is crucial to theorize the effects of each discrete negative emotion on online content diffusion separately. We customize the emotion as social information (EASI) framework \citep{van2009emotions,van2010emerging}, a generalized framework that incorporates, but is not limited to, the theories of cognitive appraisal, approach-avoidance motivation, social comparison and emotional contagion (detailed later in this section), to theorize how and why each type of negative emotional expression shapes online content diffusion, respectively.
%Recent studies have started to explore the impact of some negative discrete emotions on the retweeting of specific types of tweets, such as anger and disgust on moralized tweets \citep{brady2019ideological}, and anger and fear on cancer tweets \citep{wang2020negative}. Few studies focused on a comprehensive set of negative discrete emotional expressions on online content diffusion, and beyond the context of tweets. More importantly, there lacks a unified theoretical framework to predict and explain the effects of negative emotions on online content diffusion. 

Three theoretical frameworks of social psychology are related. The first is emotional contagion theory \citep{hatfield1993emotional,barsade2002ripple,coviello2014detecting,kramer2014experimental,del2016echo}. The theory argues that a writer's emotional state can be transferred to readers with a certain probability, through automatic mimicry and synchronization of the writer's emotions. Such synchronized emotional states can cause readers to spread the content in their social networks. This framework, however, cannot fully explain the impact of emotional expressions on online content diffusion. Emotional contagion does not necessarily occur in the context of content consumption \citep{heath2001emotional,wang2020negative}. For example, expressions of disgust could make readers feel positive due to their being able to gloat \citep{heath2001emotional}. Expressions of anger and fear embedded in tweets are not effective in arousing similar emotions in tweets' viewers \citep{wang2020negative}. Further, emotional states, such as those of low arousal, do not always activate the readers to share content with others \citep{berger2012makes}. As a result, emotional contagion theory is insufficient to explain the impact of emotional expressions on content diffusion.

The second related framework is emotion and approach-avoidance motivation (see \cite{elliot2013approach} for a review), which argues that one's approach-avoidance motivations are influenced by another person's emotional expression. For instance, individuals tend to develop an avoidance tendency toward another person's expression of anger that targets them \citep{adams2006emotional}. It is theoretically unclear, however, to directly apply this framework to theorize the effects of emotional expressions on the social transmission of online articles \citep{elliot2013approach}. In particular, writers' anger is usually expressed toward specific events rather than toward readers, and therefore, the readers are unlikely to feel threatened and develop an avoidance tendency. 

We ground our theoretical development in the third framework, the emotion as social information theory \citep{van2009emotions,van2010emerging}, which is a wildly accepted theory of the role of emotional expression in social interaction and can incorporate the other two frameworks. According to the EASI framework, a writer's emotional expression can affect readers' behavior by two mechanisms, i.e., inferential processes and affective reactions \citep{van2009emotions}. In particular, readers tend to make cognitive inferences on the content and the writer based on the writer's emotional expression \citep{keltner1999social}. Such cognitive inferences may induce an approach-avoidance intention toward the writer or the content, which speaks to the theory of \cite{elliot2013approach}. Further, the writer's emotional expression can elicit affective reactions in readers \citep{clark1991reactions}. Specifically, such affective reactions can be induced through two mechanisms. The first is emotional contagion. As we detail later, we found evidence of emotional contagion in our data. The readers tend to express the same emotions in their comments as the emotions mainly expressed by the writers in the article. Second, one's emotional expression may affect others' impressions and interpersonal liking toward the focus person \citep{van2009emotions}. For example, employees with an angry leader tend to hold a negative impression of the leader \citep{van2010emerging}. Readers' inferences and affective reactions subsequently affect their sharing behaviors. Based on the EASI framework, we discuss the effects of each negative emotion on the reader’s inferential processes and affective reactions and then the impact on content diffusion.

%Based on the framework of EASI, we outline our discussions on the effects of negative emotional expressions in Table \ref{tab:EASI}. Specifically, for each negative emotion, based on the existing literature, we discuss their effects on the reader’s inferential processes and affective reactions, respectively. Then, we predict their effects on content diffusion.

%The emotional expressions in content also can affect the audience’s emotional state and their subsequent decisions \citep{rime1991beyond,heath2001emotional,berger2012makes}. Individuals who are exposed to high-arousal emotions tend to be activated and are more likely to take action, such as sharing with peers \citep{berger2012makes}. In this regard, laboratory experiments show that urban legends that evoke a disgusting experience are more likely to be socially transmitted \citep{heath2001emotional}. 
% But there are few large-scale field investigations on whether and how emotions can effectively affect content diffusion in digital social networks and whether the effects of different emotions are heterogeneous. 

\textit{Anger.} Angry expressions tend to make the reader perceive that the author engaged in low cognitive effort \citep{yin2014anxious} and has a low level of rationality \citep{xiao2018social}, and thus the content may not be trustworthy or helpful. Such a perception may further lead the reader to develop an avoidance tendency toward the author \citep{van2009emotions}, including not engaging in socially sharing the content. Second, the reader tends not to share angry content to avoid being perceived by his or her friends as having a low cognitive capability or a low level of rationality. As for affective reactions, when encountering angry expressions, one may develop a negative impression and negative sentiment toward the author or the content \citep{van2010emerging}, which reduces the intention of sharing. Therefore, angry expressions likely affect online content diffusion negatively.

\textit{Anxiety.} Different from anger, the inferential processes of anxious expressions make the reader perceive the author to have engaged in high cognitive effort and to be prudent and rational \citep{yin2014anxious} and, thus, perceive the content as helpful and trustworthy. The reader thus can further benefit from sharing such content to make him- or herself be regarded as prudent, rational, and helpful by his or her friends. In terms of affective reactions, as anxiety is a contagious emotion \citep{heath2001emotional}, expressions of anxiety can quickly spread such anxiety to the readers \citep{hatfield1993emotional}. Readers can be activated by the feeling of anxiety and develop a strong tendency to share the content to seek advice, release their emotional burden, and gain social support \citep{rime2009emotion}. This process is also known as the ``approach to avoid'' tendency \citep{elliot2013approach}, whereby people approach social support to avoid enduring personal stress. As a result,  anxious expressions likely affect online content diffusion positively.

\textit{Sadness.} When confronted with another’s sadness, one might infer that the other faces a loss and has low coping potential \citep{van2010emerging}. The reader would not share the content publicly to avoid disclosing the other's loss and exacerbating the sadness as well as to protect one’s privacy \citep{finkenauer1998secrets}. In terms of affective reactions, sad expressions can make the reader feel sad or depressed through emotional contagion. Low arousal reduces the inclination to take action \citep{lerner2015emotion}, including reducing content sharing intention \citep{berger2012makes}. Instead, the reader would allocate more time and energy to offering help or consolation. Hence, sad expressions may negatively affect online content diffusion.

\textit{Disgust.} The reader can infer from expressions of disgust that their circumstances are better than those of the person who feels disgusted described in the article through social comparison \citep{heath2001emotional}. Regarding affective reactions, people can feel positive and activated due to gloating \citep{heath2001emotional,berger2012makes} when hearing about others' expressing disgust. Thus, the reader tends to share content with expressions of disgust because consuming them generates positive utility for both the reader and their friends \citep{heath2001emotional}. Therefore, it is likely that expressions of disgust positively affect online content diffusion.

Further, negative emotional expressions embedded in the content also may affect the characteristics of individuals and the social ties, which diffuse the articles. Older users are more susceptible to high-arousal and negative emotional expressions, such as anger and anxiety \citep{kensinger2008age}, possibly due to a ``midlife crisis'' caused by increasing age, worsening health, and a perceived lack of accomplishments \citep{wethington2000expecting}.\footnote{A ``midlife crisis'' also may be related to a feeling of sadness or depression. The middle-aged user, however, may feel reluctant to share content with expressions of sadness because the behavior could give the impression of the user's being an underdog and in need of help. Indeed, research has found that older adults are less likely to seek explicit social support than are young adults \citep{jiang2018age}.} 
In contrast, younger users are more likely to enjoy consuming other people's stories of disgust due to gloating \citep{heath2001emotional,kensinger2008age}. Females tend to share positive information more than do males \citep{lin2020examining} and, thus, may engage less in diffusing content with expressions of sadness.
Users with more contacts are of higher network centrality, which indicates higher social status and social influence. They are motivated to share content that expresses more anxiety and less anger to maintain their status and impact, as anxious expressions are associated with an image of high cognitive capability and prudence \citep{yin2014anxious}, whereas angry expressions are linked to a lack of self-control, low cognitive capability, and irrationality \citep{xiao2018social}. Further, since people have a demand of consuming stories about others having disgusting experiences, due to gloating and social comparison \citep{heath2001emotional}, weak-tie friends may share these stories to strengthen their social relationships with each other. However, people are less motivated to manage impressions with their strong-tie friends, and thus, they may feel more comfortable of sharing angry content with their close contacts. People are also likely to share with more important contacts (i.e., strong-tie friends) about the content that expresses anxiety, which is perceived as more urgent and valuable \citep{yin2014anxious}.

\section{Data}
% We collected 387,486 online articles distributed and spread on the WeChat social networking site, in which over 3.5 million active publishers generate, on average, more than 4.9 million articles per day. 

To analyze emotional expressions in online articles and their diffusion cascades, we first randomly sampled 100,000 official accounts, the publishers on WeChat,\footnote{A total of 3.5 million active publishers generate, on average, more than 4.9 million articles per day on the WeChat official account platform, a feature of the WeChat platform.} then filtered out the inactive publishers who posted fewer than 10 articles during our observation period (August 31 to November 30, 2018), and, then sampled 38,839 publishers. We recorded the average number of followers of each publisher during our observation period (publisher's popularity), the average number of articles that the publishers posted per day (publisher's proactivity), and publisher type (i.e. individual, media\footnote{Organizations with government permission to publish the content of a newspaper, radio, and television.}, business enterprises, government or other organizations).

For each of these publishers, we randomly selected 10\% of the articles that they posted during the study period and collected 387,486 online articles in total for analysis. 
This random sample well represents the articles distributed and diffused on WeChat and covers topics as diverse as politics, economics, business, society, sports, and technology, among others. The sample includes both short texts (fewer than 100 characters, similar to tweets) and long articles (more than 100,000 characters, similar to newspaper and magazine articles and chapters in a novel), with a mean article length of 1,164.60 characters and a standard deviation of 2,060.98 characters. Although the WeChat platform publishes mainly text, it allows articles to contain images and videos\footnote{Our results are robust to the sample that excluded the articles that contain mainly video but few words, which we detail in the section on robustness checks.}. We also recorded the number of images and the number of videos in each article. 

A total of 6,823,576 unique individuals shared these articles with their first-degree friends (i.e., strong-tie contacts), or their acquaintances (i.e., non-first-degree friends, weak-tie contacts)\footnote{Acquaintances (weak-tie contacts) are non-first-degree friends but are in the same group chats. Such a relationship is a ``tie'', because it allows information diffusion through the group chat. It is considered ``weak'', because these users are not direct contacts, and they are not likely to be familiar with each other.} on WeChat. We collected data on the demographic (i.e., age and gender) and network characteristics (i.e., network degree) of all of the users involved in the cascades as well as the tie strength between every pair of sender and receiver (whether they were first-degree friends) in each cascade.   

\subsection{Mapping Cascades}
\begin{figure}[H]
  \begin{center}
      \includegraphics[width=1.0\columnwidth]{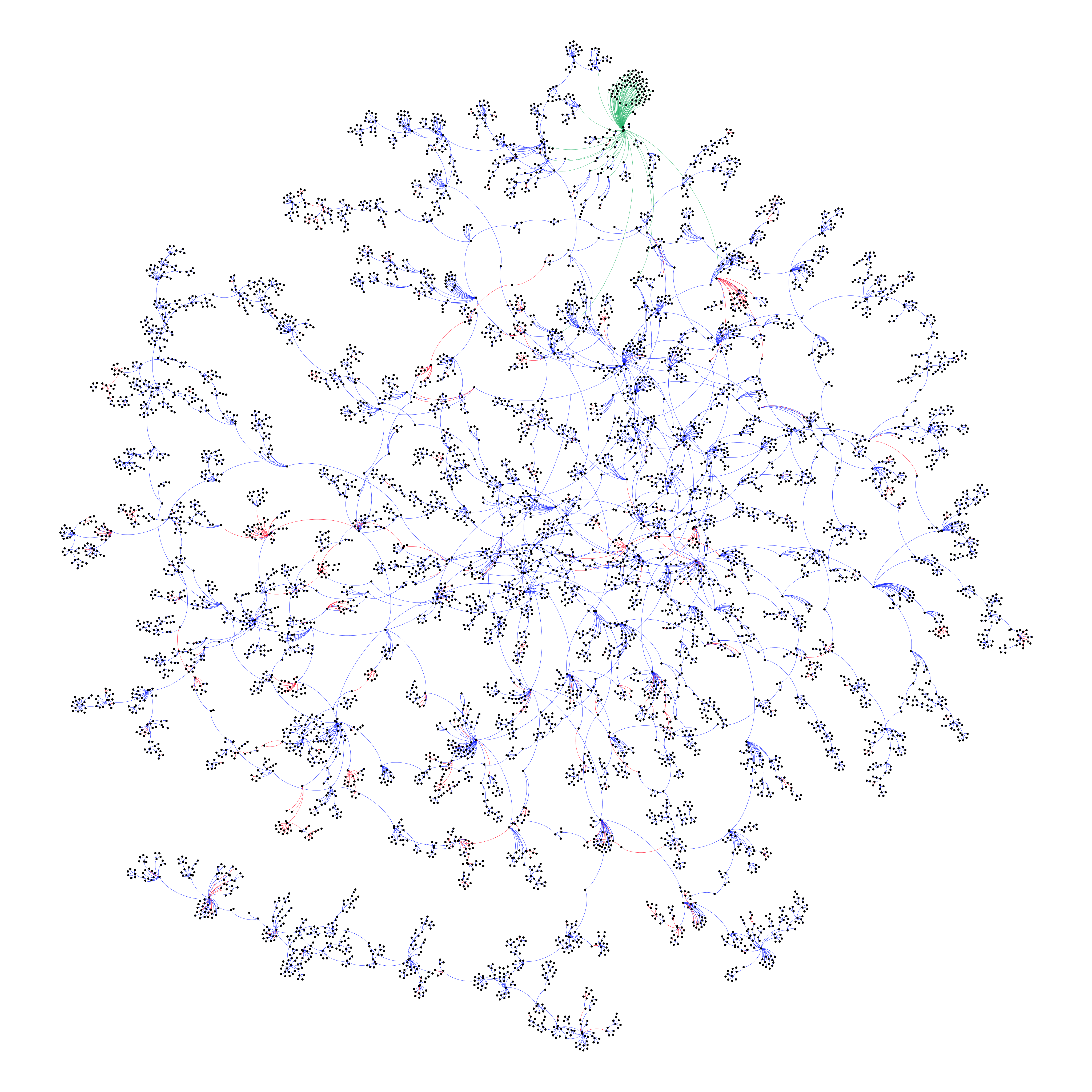}
  \end{center}
  \caption{An example of a large cascade. }
  \scriptsize Note: The cascade involves 7,225 unique individuals (represented by nodes). The green edges represent sharing from the public account. The red and blue edges represent sharing from acquaintances (weak ties) or sharing from first-degree friends (strong ties), respectively. The article documented a scandal that involved certain individual bloggers and journalists who illegally blackmailed a number of real estate firms. The bloggers and journalists threatened to spread negative online articles about the firms unless they received a large payment.
  \label{fig:Cascade}
\end{figure}

Our dataset uniquely overcomes the cascade mapping problem, which commonly occurs in Twitter retweet cascade mapping. Retweet cascades are essentially inferred based on timing and relations \citep{shi2014content,vosoughi2018spread}. Our context, however, enables us to precisely catalog the diffusion cascades of each article in our sample by recording the \textit{exact} account ID of all individuals who shared the article as well as from whom they received access to the article and the timestamp of each sharing\footnote{When a sender sends an article to a receiver, the link for the article (including article title and a short description) is displayed to the receiver. Such sharing behaviors are recorded in the platform's database. When the receiver opens the link and again shares the article, we catalog both the sender's and the receiver's ID.}. An article cascade starts when an article is posted to a publisher's followers, who then share the article in their local social networks. Users at the end of each cascade are those who shared the article, which was not shared by anyone again during the following week. On WeChat, if an article is not shared by anyone again during the following seven days, it is improbable (less than 1\% probability) that the article would be shared again. Figure \ref{fig:Cascade} shows an example of a large article cascade in our sample.

\subsection{Measuring Cascades}
% We measured the cascades in five dimensions: size (the total number of users involved in sharing the article), depth (the maximum number of sharing hops from the origin article, where a hop is a sharing action by a new unique user), maximum breadth (the maximum number of users involved in the cascade at any depth), time (the average number of hours that the cascading process takes to finish one level of depth), and structural virality (the average length of the shortest paths between all pairs of nodes in a diffusion tree) \citep{goel2015structural,vosoughi2018spread}.

As noted in Section 2.2, we measured the cascades through four dimensions: size, depth, maximum breadth, and structural virality \citep{goel2015structural,vosoughi2018spread}. These four measures are succinct representations of the shape and dynamics of a cascading process.  
Specifically, structural virality is defined as $\frac{1}{n(n-1)}\sum_{i=1}^n\sum_{j=1}^n d_{ij}$
% in Equation \ref{eq:eq1}:
% \begin{equation}
% \label{eq:eq1}
%     \begin{aligned}
%     \text{Structural virality}=\frac{1}{n(n-1)}\sum_{i=1}^n\sum_{j=1}^n d_{ij},
%     \end{aligned}
% \end{equation}
, where $n$ represents the number of individuals involved in the cascades, and $d_{ij}$ represents the shortest distance between individual $i$ and individual $j$. Provided the same level of cascade size, higher structural virality indicates that the cascade is driven more by decentralized and peer-to-peer sharing than by broadcasting. 

\begin{figure*}[t]
\begin{center}
    \includegraphics[width=0.8\columnwidth]{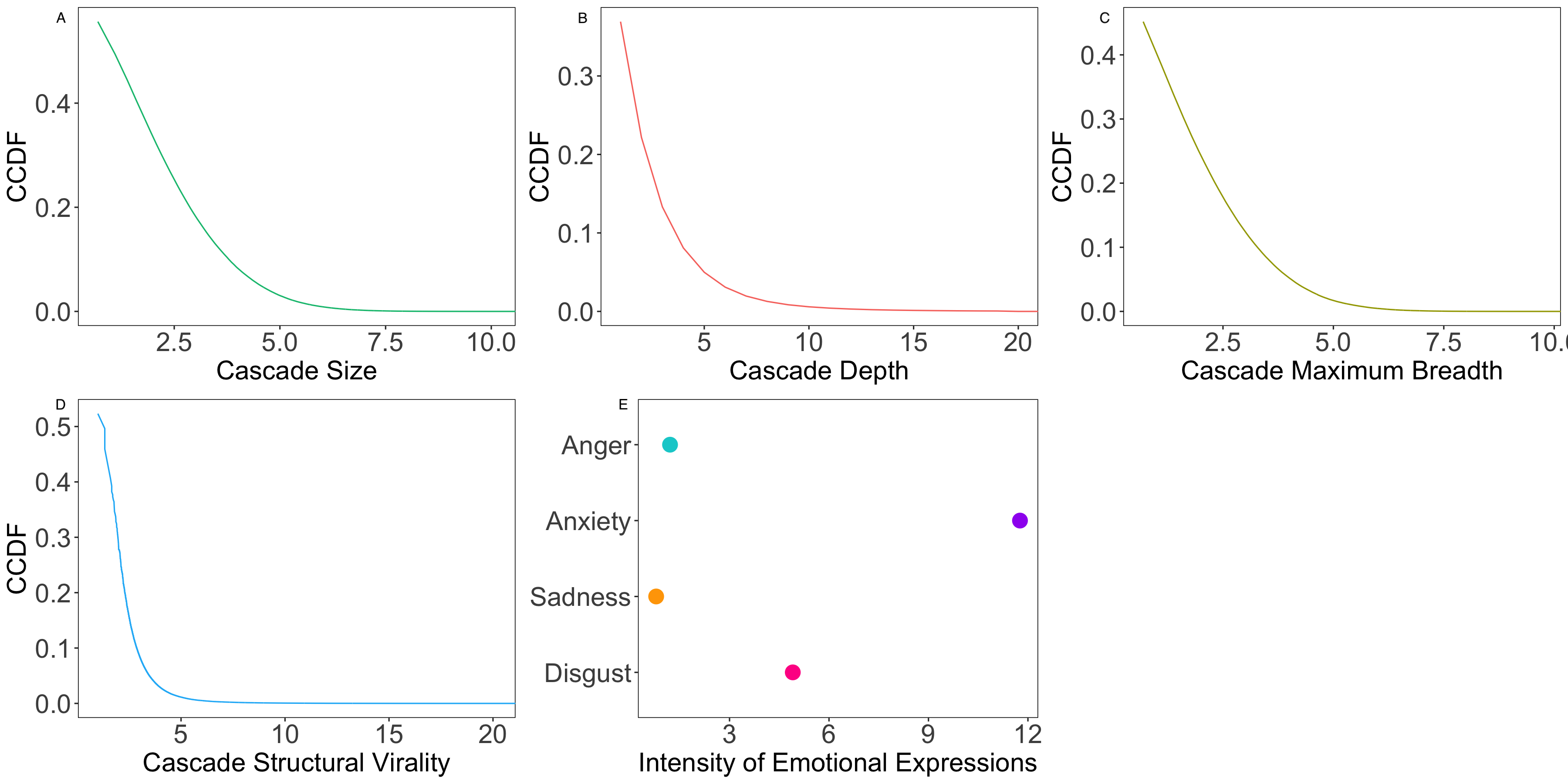}
\end{center}
\caption{Complementary cumulative distribution functions (CCDFs) of cascade scales and the level of intensity of four negative emotional expressions for an average article.} \scriptsize Note: (A-D) The CCDF of cascade size, depth, maximum breadth, and structural virality. The CCDF is the complement of the cumulative distribution function (CDF). Taking cascade size as an example: for a given level of cascade size in the x axis, the corresponding value of the CCDF shows the fraction of articles whose cascade size is above or equal to this level. (E) The level of intensity of four negative emotional expressions for an average article. The values are taken directly from our emotion detection analysis, and have not been normalized to a standard distribution (so that we can compare the mean values of different types of emotions). The differences between all pairs of the four emotions are statistically significant ($ps<0.001$).
\label{fig:ccdf}
\end{figure*}

Figure \ref{fig:ccdf} (A-D) provides a description for the empirical distributions of cascade size, depth, maximum breadth, and structural virality of the articles in our sample, and indicates that a very small fraction of the articles exhibit high values in every cascade dimension. To see the relationships among these cascades' dimensions, we present the correlation matrix (Table \ref{tab:cascadecor}). We find a strong relationship between cascades' size and maximum breadth (0.974), as 41\% of the articles' cascades ended with the first level. Another strong correlation is that between cascades' depth and structural virality (0.921). To explain the correlation, first, structural virality is the average shortest distance between each pair of the nodes in a cascade. Second, the larger depth indicates that the corresponding diagram of the cascade has a larger diameter (because cascade depth is an inferior of cascade diameter). Thus, we can expect a larger diameter to be positively correlated with the average shortest distance between each pair of nodes in the cascade. Moreover, due to the positive correlation between cascades' depth and size, structural virality also is positively correlated with cascades' size and maximum breadth. Despite the relatively high correlations, it is not trivial to include all of the dimensions as our outcome variables, because not only different dimensions have distinct implications for an understanding of diffusion, but emotions also can have disparate effects on these correlated dimensions, as we detail later.
%%%%%%%%%%%%%%%%%%%%%%%%%%%%%%%%%%%%%%%%%%%%%%%

\section{Detection of Emotional Expressions}
% We focused on discrete emotions instead of the dimensions of emotion, such as valence (positive or negative polarity of emotions) and arousal (``the extent to which a person is energized by an experience'')\citep[][p. 448]{yin2017keep}.
% The affective dimensions of content, such as embedded emotions, receive much less attention than their cognitive counterparts, due mainly to the difficulty of emotion detection in content. 
We detect word-level emotional expressions by adopting a state-of-the-art lexicon-generation approach \citep{xue2014study,yu2019emotions}. Specifically, we constructed a new domain-specific and up-to-date emotion lexicon based on a general emotion lexicon with 16,017 unique emotion words \citep{quan2010blog} and 8 million word-embedding vectors \citep{song2018directional}. We constructed a new lexicon mainly for two reasons. First, the general lexicon, constructed by using emotional expressions in Chinese blogs \citep{quan2010blog}, may miss domain-specific emotional expressions. The emotional expressions in WeChat articles can be very different from those in Chinese blogs. Second, emotional expressions are dynamic over time. The general lexicon was built in 2008, and misses the new expressions of emotion generated in the following decade.

We detect emotional expressions at the word level instead of at the document level, following the existing literature \citep{yin2014anxious,brady2017emotion,yu2019emotions}. Emotion is inherently ambiguous, subjective, and dynamic over time. It is difficult to reach an agreement on document-level emotion expression annotation \citep{quan2010blog}, which hinders training a document-level supervised learning model.\footnote{Intuitively, it is challenging, even for humans, to accurately assign a single value of anger to describe how much anger is contained in \textit{Hamlet} and whether the value should be higher for \textit{King Lear} as compared to \textit{Hamlet}.} Starting from the word level, however, mitigates document-level ambiguity. 
Moreover, although research on NLP has started to explore document-level analysis, most state-of-the-art models have a length limit of only a few hundred characters \citep{sun2019fine}, which is incompatible with the length of online articles.\footnote{For example, BERT has a length limit of 512 characters \citep{sun2019fine}. Due to the complexity of natural language, long documents will bring extreme difficulty to a model with even millions of parameters. Our sample's mean article length is 1,164.60 characters, with a standard deviation of 2,060.98 characters. Some articles can be more than 100,000 characters. This is clearly more than what state-of-the-art document-level methods can handle.} 
%Measuring word-level emotional expressions is also compatible with the theoretical framework of EASI \citep{van2009emotions,van2010emerging}.

We first use an existing emotion lexicon, Ren-CECps \citep{quan2010blog}, as a basic lexicon. Each word $w_i$ in the lexicon is mapped to an eight-dimension vector $v_i= (I^{i}_{1}, I^{i}_{2}, ..., I^{i}_{8})$, where $I^{i}_{k}\in [0,1]$ is manually annotated and represents the intensity of the $k$th discrete emotions expressed by $w_i$. Second, we retrieve word vectors that contain words' semantic information. The word vectors can be derived by statistical language modeling (e.g. Word2Vec by \cite{mikolov2013efficient}). We used pre-trained word vectors by \cite{song2018directional}, who provide 200-dimension word vectors for over 8 million common Chinese words and phrases. These word vectors are pretrained on up-to-date, large-scale, and high-quality Chinese online content, and have been validated through various NLP tasks \citep{song2018directional}. Then, the similarity between the two words can be measured by the cosine similarity of the two corresponding word vectors \citep{mikolov2013efficient}. Third, we follow the approach proposed by \cite{xue2014study} and \cite{yu2019emotions} to extend the basic lexicon to a domain-specific and up-to-date lexicon. For every emotion word in the basic lexicon, we use 8 million word vectors to mine their semantically similar words as potential emotion words. For each potential emotion word, we use its semantically nearest words in the basic lexicon to determine whether it is an emotion word and its intensities for eight types of discrete emotions. Fourth, we combine the newly mined emotion words with words in the basic lexicon to form an extended lexicon. We then treat the extended lexicon as a new ``basic lexicon'' that is used to repeat the third step until the total word amount in the combined lexicon converged (See Appendix \ref{appendix:construct} for more details). A total of 16,921 new words were found after this process, and the extended lexicon contains a total of 28,969 words. This result confirms the necessity of constructing a domain-specific lexicon, without which 58.4\% of unique emotion words (16,921 out of 28,969) would be ignored, i.e., if only the basic lexicon were used. Five raters were recruited to annotate the emotional intensities of the newly mined words for the eight discrete emotions. There is no statistically significant difference between the results generated by the algorithms and by the raters, confirming the validity of our new lexicon (See Appendix \ref{appendix:Eval}).

We summarize the emotion words in an article, and consider the negation and degree words (if any) associated with these emotion words.
%\footnote{Every emotion word expresses a few types of discrete emotions at certain intensity levels. In our lexicon, each emotion word has eight intensity values (from 0.0 to 1.0), each of which is associated with one type of discrete emotions. For each article, our algorithm first looks for the emotion words that are contained in the lexicon. Second, if our algorithm finds such an emotion word, it will then check and capture any negation and degree words that are associated with the emotion word. Third, we adjusted the emotional intensities of the emotion word by considering these degree words and negation words. Finally, for each type of discrete emotions, we sum up all the adjusted intensities of all emotion words that appear in the article, as the level of the emotional intensity of the article. This approach is better than merely counting the number of emotion words, because we treat the emotion intensity as continuous values, and we consider the effects of degree words and negation words on emotion expression.} (See Section \ref{appendix:emo} in the Appendix). 
In accordance with the document-level emotion expression space model \citep{quan2010blog}, we map each online article into an eight-dimensional vector. Each element of the vector represents the emotional intensity of its corresponding discrete emotion expressed in the article $d$: $d=(e_1,...,e_k,...,e_8),$ where $e_k$ is determined by emotion, negation, and degree words contained in the article. Negation and degree words are frequently used in Chinese and thus are helpful for accurate emotion analysis \citep{quan2010blog}. We adopted a negation word dictionary provided by TextMind, a Chinese language psychological analysis system developed by the Chinese Academy of Sciences. The dictionary contains 31 frequently used negation Chinese words. We used 60 degree words provided by Ren-CECps as our degree-word dictionary and annotated their degree values. For example, the degree value of the word meaning ``the most'' in Chinese is annotated as 1.5, and that of the word meaning ``kind of'' is annotated as 0.8.

We use a forward-sliding window (with the window size set to three words) to capture these negations and degree words. If our algorithm finds an emotion word in the article, it checks the three words before the emotion words and captures any negation and degree words that appear within these three words. We choose three as sliding-window size because normally in Chinese, negative and degree words are within three words before the corresponding emotion words. The $k$th discrete emotion intensity of the article $d$, denoted as $e_k(d)$, is determined as follows:

\begin{equation}
    \begin{aligned}
    e_k(d)=\Sigma_{i=1}^{n} (-1)^{m_i}\times DegV_i\times I_k(w_i),
    \end{aligned}
\end{equation}
where $\{w_i\}_{i=1}^{n}$ are the emotion words in both article $d$ and our lexicon. It expresses the $k$th discrete emotion.  If $n$ = 0, then $e_k (d)$  will be set to zero. $I_k (w_i)$ refers to the $k$th discrete emotion intensity of $w_i$ $(k\in\{1,2,…,8\})$. $m_i$ is the total number of negative words that appear in the sliding window of $w_i$. Finally, $DegV_i$ is the average degree value of all degree words that appear in the sliding window of $w_i$. The emotions of each article are then summarized into an eight-dimensional vector. We used the same procedure and analyzed eight discrete emotions in the comments. 

We present the average intensities of emotional expressions on the article-level. We find that among all of the negative emotions, anxiety has the highest level of emotion intensity (See Figure \ref{fig:ccdf} (E)). Further, we present the correlation matrix of the four discrete negative emotions in Table \ref{tab:emocor} to check their independence. The correlations of all pairs of negative emotions are below 0.44, whereas most of the correlations are below 0.20. The result implies that the correlations among the four negative emotions are low, indicating the independence of these four negative emotions.

These raw discrete emotion intensities (elements in the eight-dimensional vectors) are of different scales (i.e., with different population means and variances). To enable the comparison of different discrete emotions under the same scale, we standardize these raw discrete emotion variables to z-scores. A z-score is calculated by subtracting the population mean from an individual raw score and then dividing the difference by the population standard deviation. The standardization makes the intensities of discrete emotions fluctuate around a zero mean and measured on a scale of 1 standard deviation (close to the standard normal distribution) \citep{bollen2011twitter}. Thus, if an article has a 0.0 intensity score in anxiety, it would contain more expressions of anxiety (emotion words and relevant degree words) than about 50\% of the articles in our sample; 1.0 means more than about 84\%; 2.0 means more than about 97.5\%, and so forth. For brevity, hereafter, we refer the normalized intensity scores as intensity score.

\section{Model}
\subsection{Instruments of Emotional Expressions}
Intensity scores of emotional expressions embedded in articles are likely to be endogenous to content diffusion outcomes. For example, articles on natural disasters may generally contain a higher degree of anxiety, and tend to be widely shared. To address the endogeneity concern and inspect the causal relationship between emotional expression and information diffusion, we applied an instrumental variable approach, commonly used in estimating the effects of online content characteristics \citep{dev2019quantifying}. In our model, we used two sets of instrumental variables: intensity scores of the emotional expressions embedded in the latest article right before the current article of the current publisher (i.e., lagged emotional expression intensity scores)\footnote{One may concern that the time difference between the latest article and the current article published by the same publisher may not be long enough, and therefore the publisher may still face the same environment when publishing the latest and current articles (as an unobservable confounding factor in the model identification). We find, however, that, on average, a publisher would not post a new article until 9.295 days passed from publishing her latest article. Publishers were, thus, unlikely to face the same environment when publishing the latest and current articles. We choose only to use intensity scores of the latest article as instruments, not to use those of more historical articles, because if the time lag is too long, the correlation between intensity scores of historical articles and the current article would be weak, which could lead to the weak instrument problem.} and average intensity scores of the emotional expressions embedded in the articles published by other publishers in the same period as the current article\footnote{Articles of the same period refer to articles published neither earlier nor later for more than seven days as compared to the current article.}.
% (e.g., exogenous events which happen in a specific period, endogenous decision process of a publisher when publishing a new article, and measurement errors of intensity scores)

Instrumental variable estimators are also well established to correct for measurement errors \citep{Wooldridge2002}. Our two sets of instrumental variables can be applied to further mitigate the concern of measurement errors in emotional expression intensity scores, although measurement errors should not be a significant issue in our identification. Appendix \ref{appendix:Eval} already shows that the measurement generated by the algorithm is not statistically significantly different from that developed by human coders. Given that the measurement errors are very unlikely to be systematic across articles, the measurement errors in intensity scores of the focal article are not likely to be correlated with our two sets of instrumental variables.\footnote{Measurement errors might rise in some extreme cases, such as when our lexicon fails to capture a very uncommonly used emotion word, or when in a special context, the intensity of an emotion word deviates greatly from that in our lexicon. But if these cases do not systematically happen, the measurement errors cannot bias our estimation. Indeed, WeChat articles, published by different publishers or published by the same publisher at different time points, contain a large variety of topics and contexts. The emotional expressions associated with these topics and contexts are thus of wide variety. Indeed, we find at least 28,969 unique emotion words used in our sample. Therefore, we do not expect a measurement error associated with a particular emotional expression to be systematically contained in a large number of articles. Instead, we believe it is reasonable to assume the measurement errors as independent across articles.} Due to this orthogonality, the measurement errors would not bias the estimation results \citep{Wooldridge2002,chernozhukov2018double}.
%Despite their correlation with the emotional expression intensity scores of the current article, the same measurement error is unlikely to exist in the previous articles of the same publisher or in the concurrent articles published by other publishers.

In Section \ref{sec:insttest}, we ran the weak instrument test \citep{16284} and Sargan test \citep{10.2307/1907619}, which confirmed the validity (i.e., relevance and exogeneity) of our instruments. Further, we implemented a double machine learning framework \citep{chernozhukov2018double} to estimate a partial-linear instrumental variable specification. This specification enables us to control the non-linear effects of covariates with machine learning techniques and to keep the relationship between the outcomes (i.e., cascade characteristics) and variables of interest (i.e., emotion intensity scores) linear for interpretability. Next, we present our two sets of instrumental variables. 

Lagged independent variables are commonly used as instrumental variables in economic and marketing research \citep{villas1999endogeneity}. We therefore propose our first set of instrumental variables, lagged emotional expression intensity scores, to address possible endogeneity problems. Intensity scores of the emotional expressions embedded in the latest article (i.e., the article right before the current article) posted by the current publisher can be valid instruments, because they tend to have a correlation with emotional expressions in the current article, in that the publisher's emotional states can pervasively carry over from one situation to next \citep{lerner2004heart,lerner2015emotion}. However, lagged emotional expression scores cannot directly affect the diffusion of the current article. The only way that they may correlate with the diffusion of the current article is through correlating with the emotional expression in the current article.
% when composing an article, a publisher's decision on content may be affected by some exogenous factors in the current period, which may also impact the article's diffusion. For example, articles about natural disasters may generally contain a higher degree of anxiety, and tend to be widely shared. To address this endogeneity problem, we propose our first set of instrumental variables, lagged emotional expression intensity scores. Intensity scores of the emotional expressions embedded in the latest article (i.e., the article published right before the current article)
% posted by the current publisher right before the current article can be valid instruments for emotional expression intensity scores in the current article because both articles are posted by the same publisher and therefore the corresponding emotional expression intensity scores are correlated due to publisher-specific habits and writing styles. However, lagged emotional expression intensity scores are uncorrelated with the random shocks faced by current publisher when publishing the current article because of time difference, and emotional intensity scores are also unlikely to be serially correlated after controlling publisher-fixed effects. 

% \subsubsection*{Average emotional expression intensity scores of concurrent articles by other publishers:} 
Following \cite{nevo2000mergers} and \cite{hausman20085}, we propose our second set of instrumental variables, average emotional expression intensity scores of concurrent articles by other publishers. This set of variables are valid instruments. On one hand, they tend to correlate with the emotional expression in the current article, because it is well documented that in social media, individuals' emotional expressions tend to be synchronized due to emotional contagion \citep[e.g.,][]{hatfield1993emotional,barsade2002ripple,del2016echo}. On the other hand, the emotional interactions among publishers can hardly be observed by readers. The readers' decision of whether to share the article only depends on the content (including emotional expression) in the current article. 
%the publishers in our dataset are randomly sampled from a large pool of publishers, and their audiences are unlikely to be highly overlapping. It is thus implausible that emotional expressions embedded in articles published by other publishers could be received by the current publisher's audience and then directly affect the cascades of current publisher's articles. 
Therefore, we argue that average emotional expression intensity scores of concurrent articles by other publishers cannot directly affect the diffusion outcomes of current publisher's articles. This set of instrumental variables may
correlate with the diffusion outcomes of current publisher's articles only through affecting the current publisher's emotional expression in the articles.

\subsection{Effects of Negative Emotional Expressions on Structural Properties of Cascades}
We implement a partial-linear instrumental variable approach to estimate the effects of emotional expressions on cascade dimensions. The model is shown as follows:
\begin{subequations}
\begin{align}
y_{ij}&=\bm{\beta^{'}}\mathbf{emotion}_{ij}+g_0(\mathbf{article}_{ij})+Publisher_i+\epsilon_{ij},\\
\mathbf{emotion}_{ij}&=\bm{\beta_1^{'}}\mathbf{z}_{ij}+h_0(\mathbf{article}_{ij})+\bm{\eta}_{ij},\\
\mathbf{z}_{ij}&=m_{0}(\mathbf{article}_{ij})+\bm{\mu}_{ij},
\end{align}
\label{eq:de_main}
\end{subequations}
where $y_{ij}$ indicates one of the four cascades’ structural properties of the $jth$ article published by the $ith$ publisher, $\mathbf{emotion}_{ij}$ indicates the intensity scores of (discrete) emotions of the $jth$ article published by the $ith$ publisher, $\mathbf{article}_{ij}$ indicates the article-level control variables of the $jth$ article published by the $ith$ publisher (i.e., article length, number of images and videos embedded in the article, whether the article was posted during a weekend, the number of comments, and the topic distribution of the article), $Publisher_i$ indicates the publisher fixed effects, and $\mathbf{z}_{ij}$ indicates the instrumental variables that we incorporate (i.e., lagged intensity scores of emotions and average intensity scores of emotions in concurrent articles published by other publishers) which may affect cascades' structural properties only through shifting intensity scores of emotions in the $jth$ article published by the $ith$ publisher, conditional on article characteristics. We use $\bm{\beta}$ to capture the effect of emotion, non-linear function $g_0$ to capture how article characteristics affect cascades' dimensions, non-linear function $h_0$ to capture how article characteristics affect emotional expression intensity scores, and non-linear function $m_0$ to capture how article characteristics affect instrumental variables. $\epsilon_{ij}$, $\bm{\eta}_{ij}$, and $\bm{\mu}_{ij}$ are stochastic errors. $\mathbf{z}_{ij}$ can only shift $y_{ij}$ through affecting $\mathbf{emotion}_{ij}$, conditional on $\mathbf{article}_{ij}$, but is uncorrelated with $\epsilon_{ij}$.

To control for the article characteristics, we first estimated a latent Dirichlet allocation (LDA) model \citep{blei2003latent} based on all sampled articles and controlled for the latent topic distribution of an article by a 30-dimension vector (see Appendix \ref{appendix:wechat} and Figure \ref{fig:LDA} for more details.). We also included article length (log-transformed number of characters), number of images and videos embedded in the article (media richness), whether the article was posted during a weekend (time effect), and the number of comments in our model. To capture time-invariant factors that may have an impact on information diffusion, such as publisher specific characteristics, we include publisher fixed effects. 

To estimate our partial-linear instrumental variable approach, we applied a double machine learning framework following the procedures developed by \cite{chernozhukov2018double}. To be more specific, we first rewrite the model specified in Equation \ref{eq:de_main} in the following residual form:
\begin{subequations}
\begin{align}
w_{ij}&=\bm{\beta^{'}}\mathbf{v}_{ij}+Publisher_i+\epsilon_{ij},\\
\mathbf{v}_{ij}&=\bm{\beta_1^{'}}\bm{\mu}_{ij}+\bm{\eta}_{ij},\\
w_{ij}=y_{ij}-k_0(\mathbf{article}_{ij}),\ \mathbf{v}_{ij}&=\mathbf{emotion}_{ij}-l_0(\mathbf{article}_{ij}),\ \bm{\mu}_{ij}=\mathbf{z}_{ij}-m_{0}(\mathbf{article}_{ij}).
\end{align}
\label{eq:de_main2}
\end{subequations}
The double machine learning framework then proceeds as follows. First, we estimate non-linear functions $k_0$, $l_0$, and $m_0$ as described in Equation \ref{eq:de_main2} by $\hat{k}_0$, $\hat{l}_0$, and $\hat{m}_0$, which amounts to predicting $y_{ij}$, $\mathbf{emotion}_{ij}$, and $z_{ij}$ using $\mathbf{article}_{ij}$ with machine learning techniques. Naively training a machine learning model to fit $k_0$, $l_0$, and $m_0$, however, could result in very high bias in finite-sample estimates because the estimator would generally have a slower than $1/\sqrt{n}$ rate of convergence. One insight derived by \cite{chernozhukov2018double} is that we can use cross-fitting estimators to address this issue. In particular, we group the data at the publisher level and divide the data into $D$ evenly-sized folds that contain different publishers. For each fold $d=1,2,...,D$, we run a machine learning model on the other $D-1$ data folds to estimate the functions $k_0$, $l_0$, and $m_0$. The estimated functions are denoted by $\hat{k}_0^{(-d(i))}$, $\hat{l}_0^{(-d(i))}$, and $\hat{m}_0^{(-d(i))}$, where $d(i)\in \{1,2,...,D\}$ denotes that the fold that contains the $ith$ publisher, and $-d(i)$ indicates that we exclude fold $d(i)$ when we train the machine learning model and generate the predicted value for the $ith$ publisher. Then we have estimated residuals $\hat{w}_{ij}=y_{ij}-\hat{k}_0^{(-d(i))}(\mathbf{article}_{ij})$,
$\hat{\mathbf{v}}_{ij}=\mathbf{emotion}_{ij}-\hat{l}_0^{(-d(i))}(\mathbf{article}_{ij})$, and $\hat{\bm{\mu}}_{ij}=\mathbf{z}_{ij}-\hat{m}_{0}^{(-d(i))}(\mathbf{article}_{ij})$. We choose XGBoost as the machine learning algorithm in our model, which is commonly considered to be a state-of-the-art algorithm for many machine learning tasks and has been wildly adopted by data scientists and researchers \citep{chen2016xgboost}. We choose $D=10$, following the practice of \cite{chernozhukov2018double}.\footnote{We also ran the same model with 20-fold cross-fitting estimators and the results are consistent.} Finally, we estimate $\bm{\beta}$ by performing Two-Stage Least Squares (2SLS) regression on the estimated residuals. Specifically, we first regress $\hat{\mathbf{v}}_{ij}$ on $\hat{\bm{\mu}}_{ij}$, and obtain the predicted values $\hat{\mathbf{v}}_{ij}^{pred}$. We then estimate $\bm{\beta}$ by regressing $\hat{w}_{ij}$ on $\hat{\mathbf{v}}_{ij}^{pred}$ with controlling for publisher fixed effects. 
%The results give us estimated residuals $\hat{w}_{ij}=y_{ij}-\hat{k}_0(\mathbf{article}_{ij})$,
%$\hat{v}_{ij}=emotion_{ij}-\hat{l}_0(\mathbf{article}_{ij})$, and %$\hat{\mu}_{ij}=z_{ij}-\hat{m}_{0}(\mathbf{article}_{ij})$. 
%To avoid biases from overfitting, we also implement 10-fold cross validation

\subsection{Effects of Negative Emotional Expressions on Demographics and Social Ties of Cascades}
Based on the main model, we further investigated the effects of negative emotional expressions on the demographics (i.e., age, gender, and network degree) of all the users involved in the cascades and the social ties among them. As we detail in Section \ref{section:mech}, we tested the effects of negative emotional expressions on the average age, gender composition, average network degree (measured by users' average number of friends), and social ties (measured by the proportion of weak ties) involved in a cascade, respectively. We utilize the same partial-linear instrumental variable approach for analysis, where $y_{ij}$ in Equation \ref{eq:de_main} in this analysis indicates user demographics and their social ties involved in the cascade of the $jth$ article published by the $ith$ publisher. Similar to our analysis of the effects of negative emotional expressions on structural properties of cascades, we also controlled for the positive emotional expressions embedded in the articles.
%(i.e., average age, gender composition, average network centrality, and proportion of weak ties).
%The result shows that sadness, anxiety and love are significant determinants of users’ age (Table EC.1 A1); anxiety and love are significant determinants of users’ average friend number; sadness and love are significant determinants of weak tie proportion among users; sadness and anxiety are significant determinants of the proportion of first-level users’ who are friends with each other. 

\subsection{Instrumental Variable Test}\label{sec:insttest}
To address the concern about the validity of our instruments, we implemented weak instrument test \citep{16284} and Sargan test \citep{10.2307/1907619} to examine the strength and exogeneity of the instruments. The p-values of the F-statistics in weak instrument test are all smaller than 0.001, which indicates that all of our instruments are strong instruments. Moreover, the p-values of the J-statistics in Sargan test are all larger than 0.950, which indicates that we cannot reject the hypothesis that our instrumental variables (i.e., lagged emotional intensity scores and average emotional expression intensity scores of concurrent articles published by other publishers) are exogenous. 
%Similarly, provided that the second set of instrumental variables is exogenous, we cannot reject the exogeneity of the first set of instrumental variables.
%The results imply that our model is not overidentified in that given the exogeneity of our first set of instruments, our second set of instruments are also uncorrelated with the random shocks.

\section{Results}
% \subsection{Emotional Expressions in Articles}
% These raw discrete emotion intensities are of different scales (i.e., with different population means and variances). To enable the comparison of different discrete emotions under the same scale, we standardize these raw discrete emotion variables to z-scores. A z-score is calculated by subtracting the population mean from an individual raw score and then dividing the difference by the population standard deviation. The standardization makes the intensities of discrete emotions fluctuate around a zero mean and measured on a scale of 1 standard deviation (close to the standard normal distribution) \citep{bollen2011twitter}. Thus, if an article has a 0.0 intensity score in anxiety, it would contain more expressions of anxiety (emotion words and relevant degree words) than about 50\% of the articles in our sample; 1.0 means more than about 84\%; 2.0 means more than about 97.5\%, and so forth. For brevity, hereafter, we refer the normalized intensity scores as intensity score.

\subsection{Effects of Negative Emotional Expressions on Structural Properties of Cascades}
\begin{figure*}[t]
\begin{center}
    \includegraphics[width=0.6\columnwidth]{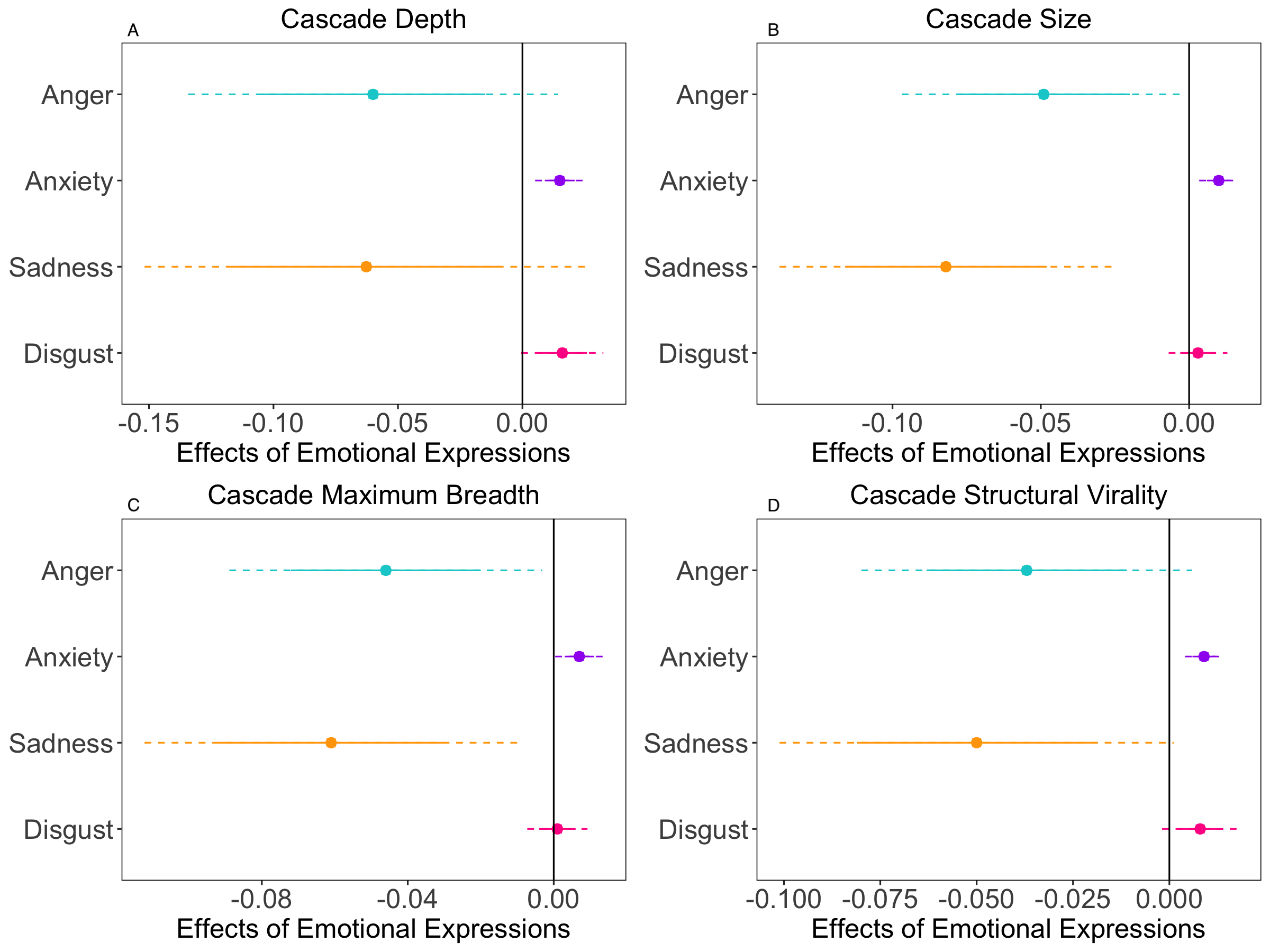}
\end{center}

\caption{The effects of four negative emotional expressions embedded in content on article cascade (A) Depth, (B) Size, (C) Maximum breadth, (D) Structural virality.}
\scriptsize Note: The $x$ axis displays the size of model coefficients. The $y$ axis displays four negative emotional expressions. The squares in the middle of the intervals represent mean values. The intervals covered by dashed lines represent 90\% confidence intervals. The intervals covered by solid lines represent mean values plus or minus 1 unit of their standard deviations. 
\label{fig:disEmo}
\end{figure*}
%%%%%%%%%%%%%%%%%%%%%%%%%%%%%%%%%%%%%%%%%%%%%%%%%%%%%%%%%%%%%%

We implemented a partial-linear instrumental variable approach to investigate the relationship between the negative emotional expressions embedded in content and the structural properties of the articles' cascade,  following the double machine learning framework as detailed in the previous section.
% To be more specific, we first applied the Frisch-Waugh-Robinson style approach to extract the effects of the article-level characteristics on each cascade dimension, emotions embedded in the articles, and the instrumental variables (i.e., lagged emotional expression intensity scores and average emotional expression intensity scores in articles published by other publishers in the same period). We used XGBoost algorithm \citep{chen2016xgboost} for prediction and applied 10-fold cross validation to avoid biases resulting from overfitting\footnote{We ran the same model with 20-fold cross validation and the results are consistent.}. We then applied the instrumental variable estimator to the orthogonalized residuals from the previous step. 
The results are shown in Table \ref{tab:tablet1} and Figure \ref{fig:disEmo}. As noted in Section \ref{lr:cv}, we focus on the effects of four negative emotional expressions (i.e., anger, anxiety, disgust, and sadness), with controlling for the effects of positive emotional expressions (See Appendix \ref{appendix: emopos_main} for details). Figure \ref{fig:disEmo} shows that three negative emotional expressions (i.e., anger, anxiety, and sadness) play significant roles in article diffusion. Articles with a higher degree of anger and sadness embedded in content exhibit a significantly lower expected cascade size ($\beta_{anger}^{size}=-0.049$, $p<0.05$; $\beta_{sadness}^{size}=-0.082$, $p<0.01$), and breadth ($\beta_{anger}^{breadth}=-0.046$, $p<0.05$; $\beta_{sadness}^{breadth}=-0.061$, $p<0.01$). In contrast, those with a higher degree of anxiety will lead to significantly larger expected cascades' depth ($\beta_{anxiety}^{depth}=0.015$, $p<0.01$), size ($\beta_{anxiety}^{size}=0.010$, $p<0.01$), breadth ($\beta_{anxiety}^{breadth}=0.007$, $p<0.01$), and structural virality ($\beta_{anxiety}^{sv}=0.009$, $p<0.01$). Disgust, however, does not exhibit a significant effect on any cascade dimension. 
These results suggest that a 1 unit (1-standard-deviation) of intensity score increase in anxiety can lead to, on average, a 1\% increase in size, 0.7\% increase in breadth\footnote{The coefficients of size and maximum breadth are interpreted as percentages, as we used log-transformation on the two dependent variables due to their right-skewed distributions.}, a 0.015-unit increase in depth, and a 0.009-unit increase in structural virality (indicating that the cascade contains more peer-to-peer sharing and fewer broadcasting structures). 
% In our sample, roughly every 100 anxiety words can make the article's diffusion cascades 4.2\% larger, 2.9\% broader, 0.063 levels deeper, and 0.038 more viral\footnote{In our sample, the maximum anxiety intensity score is 1455 units, which can make the article's diffusion cascades 1455.0\% larger, 1018.5\% broader, 21.825 levels deeper, and 13.095 more viral.}.
In addition, a 1 unit of intensity score increase in anger can result in, on average, a 4.9\% decrease in size as well as a 4.6\% decrease in maximum breadth, and a 1 unit more sadness can lead to an 8.2\% decrease in size and a 6.1\% decrease in maximum breadth\footnote{A 1 unit of intensity score of anxiety roughly consists of 24.43 anxiety words, a 1 unit of intensity score of angry roughly consists of 3.60 angry words, and a 1 unit of intensity score of sadness roughly consists of 2.59 sadness words.}. These results are consistent with the observation that anxiety is the most-expressed but anger and sadness are the least-expressed negative emotions in our sample of online content (See Figure \ref{fig:ccdf} (E)). On average, the intensity score of anxiety is 11.724, whereas the intensity scores of anger and sadness are 1.212 and 0.790. 
% Every 100 anger words can roughly lead to cascades 140\% smaller and 130\% narrower, while every 100 sadness words can result in 320\% smaller and 240\% narrower article cascades\footnote{The maximum intensity score of anger in our sample is 210 units, which can lead to cascades 1029\% smaller and 966\% narrower, while the maximum sadness intensity score is 141 units, which can result in 1156.2\% smaller and 860.1\% narrower article cascades.}.
%Both anger and sadness exhibit significant negative effects on content spread in size and breadth without affecting depth and structural virality, after holding constant the article-level characteristics and publisher fixed effects. 
% Articles with a higher degree of sadness and surprise exhibit a faster expected cascade speed ($\beta_{sadness}^{time}=-0.175$, $p<0.001$; $\beta_{surprise}^{time}=-0.176$, $p<0.01$), whereas those with a higher degree of love are associated with a slower expected cascade speed ($\beta_{love}^{time}=0.347$, $p<0.001$). Articles with a higher degree of anticipation are associated with a larger expected maximum breadth ($\beta_{anticipation}^{breadth}=0.012$, $p<0.01$), whereas those with a larger degree of joy exhibit smaller expected maximum breadth ($\beta_{joy}^{breadth}=-0.013$, $p<0.01$).

Our results are consistent with the theoretical predictions that we derive based on the EASI framework \citep{van2009emotions}, presented in Section \ref{lr:cv}. The results do not support the theory of arousal increasing social transmission \citep{berger2012makes}, as we demonstrate that the high-arousal emotion anger significantly and negatively affects cascade size and breadth. The effect of disgust is positive but insignificant, possibly because the effect of disgust is heterogeneous across different types of content and user groups. Specifically, research on the diffusion of urban legends shows that the effect of disgust is significantly positive \citep{heath2001emotional}, but research on news articles shows that the effect is insignificant \citep{berger2012makes}. Further, as we detail in the next section, content with expressions of disgust is shared by younger users and users with more friends, which also could potentially make the effect of disgust heterogeneous, and on average makes it insignificant. Moreover, as we detail in robustness checks, when we exclude articles with videos and few words and control for information originality, the effect of disgust is significantly positive on cascade depth and structural virality. 

\begin{table}
\centering
\caption{\footnotesize Main Results: Emotional Expressions\tnote{1}}
   \label{tab:tablet1}
 \begin{adjustbox}{width={0.8\textwidth},totalheight={1.0\textheight},keepaspectratio}
 \begin{threeparttable}
  \begin{tabular}{l*{5}{D{.}{.}{-1}}}
  %\begin{tabularx}{\textwidth}{@{} l
  %*{5}{S}}
    \toprule
    \multicolumn{1}{c}{\textbf{ } }    & \multicolumn{1}{c}{\textbf{Depth}}     & \multicolumn{1}{c}{\textbf{Size\tnote{2}}} & \multicolumn{1}{c}{\textbf{Breadth\tnote{2}}} & \multicolumn{1}{c}{\textbf{SV\tnote{5}}} \\
    \midrule
    \multicolumn{1}{l}{\textbf{Variable}}\\
    \cmidrule(r){1-5}
    Anger&-0.060&     -0.049*&  -0.046*&   -0.037	    \\
         & (0.045)\tnote{6} & (0.029)&  (0.026)&  (0.026)\\
    Anxiety&0.015**&    0.010**&  0.007**&  0.009**\\
           &(0.006) &   (0.004) & (0.004) & (0.003) \\
    Sadness&-0.0627  &   -0.082** &-0.061**&  -0.050\\
           &(0.054)&    (0.034) & (0.031) & (0.031) \\
    Disgust&0.016  &    0.003 &   0.001 &   0.008  	   \\
           &(0.010) &   (0.006) & (0.005) & (0.006) \\
    % Joy&-0.004&     -0.003* &  -0.002 &  -0.002\\
    %   &(0.003) &   (0.002)&  (0.002)&  (0.002)\\
    % Love&0.006*** &   0.004*** & 0.002* & 0.004***\\
    %     &(0.002)  &  (0.001) & (0.001)&  (0.001) \\
    % Surprise&-0.009  &    0.009  & 0.017** &  -0.004 \\
    %         &(0.011)  &  (0.008)&  (0.008) & (0.007) \\
    % Anticipation&0.004 &     -0.003 &  -0.004 &  -0.001 \\
    %             & (0.005) &   (0.003) & (0.002)&  (0.003) \\
    \cmidrule(r){1-5}
    % \multicolumn{1}{l}{\textbf{Article-level Controls\tnote{3}}}\\
    % \cmidrule(r){1-6}
    % %Originality&0.093***&0.06***&0.051***&-0.052&0.056***\\
    % %Info. Uniqueness&-0.057***&-0.02***&0.021***&0.079&-0.032***\\
    % Article Length&0.008&0.034***&0.041***&-0.294***&-0.004\\
    % \# of Images&0.09***&0.042***&0.035***&-0.095*&0.05***\\
    % \# of Videos&0.163***&0.103***&0.096***&0.296***&0.081***\\
    % Post at Weekends&-0.004&0.006***&0.008***&0.064*&-0.001\\
    % \# of Comments&0.218***&0.114***&0.099***&-0.003&0.101***\\
    % Reward Function&0.03***&0.029***&0.028***&-0.106*&0.02***\\					
    % \cmidrule(r){1-6}
    % \multicolumn{1}{l}{\textbf{Publisher-level Controls}}\\
    % \cmidrule(r){1-6}
    % Ave. \# of Posts&-0.195***&0.035***&0.109***&-0.597***&-0.084***\\
    % \# of Followers&0.35***&0.575***&0.643***&1.484***&0.132***\\
    % Type: Individual&-0.009&0.016*&0.022**&0.252**&-0.005\\
    % Type: Business&-0.123***&-0.087***&-0.079***&0.348*&-0.089***\\
    % Type: Gov.&-0.066***&-0.022&-0.008&0.704***&-0.05***\\
    % Type: Media&-0.096*&-0.075&-0.055&0.449&-0.066**\\
    % \cmidrule(r){1-6}
    %\hline
    \textbf{Positive emotional expressions\tnote{3}}& \checkmark & \checkmark & \checkmark & \checkmark\cr
    \textbf{Article-level characteristics\tnote{4}}& \checkmark & \checkmark & \checkmark & \checkmark\cr
    \textbf{Publisher fixed effects}& \checkmark & \checkmark & \checkmark & \checkmark\cr
    \bottomrule
  \end{tabular}
  \scriptsize Note: (1)\textit{Significance level: * $p < 0.05$, ** $p<0.01$, *** $p < 0.001$} (2) \textit{Given the right-skewed distributions of size and maximum breadth, we did a log-transformation on these two variables.} (3) \textit{Positive emotional expressions, i.e., joy, love, surprise, and anticipation, are controlled.} (4) \textit{All of the article-level variables(i.e. article length, number of images and videos embedded in the article, whether the articles was posted during a weekend, number of comments and topic distribution of the article) are controlled.} (5) \textit{SV indicates structural virality.} (6) \textit{Standard error of estimation.}
%   \begin{tablenotes}
%   %\small
%   \scriptsize
%     \item[1] \textit{Significance level: * $p < 0.05$, ** $p<0.01$, *** $p < 0.001$}
%     \item[2] \textit{Given the right-skewed distributions of size and maximum breadth, we did a log-transformation on these two variables.}
%     \item[3] \textit{Positive emotional expressions, i.e., joy, love, surprise, and anticipation, are controlled.}
%     \item[4] \textit{All of the article-level variables(i.e. article length, number of images and videos embedded in the article, whether the articles was posted during a weekend, number of comments and topic distribution of the article) are controlled.}
%     \item[5] \textit{SV indicates structural virality.}
%     \item[6] \textit{Standard error of estimation.}
%     \end{tablenotes}
\end{threeparttable}
\end{adjustbox}
\end{table}
%\end{landscape}
\subsection{Effects of Negative Emotional Expression on Demographics and Social Ties of Cascades}
\label{section:mech}
\begin{figure*}[htbp]
\begin{center}
    \includegraphics[width=0.6\columnwidth]{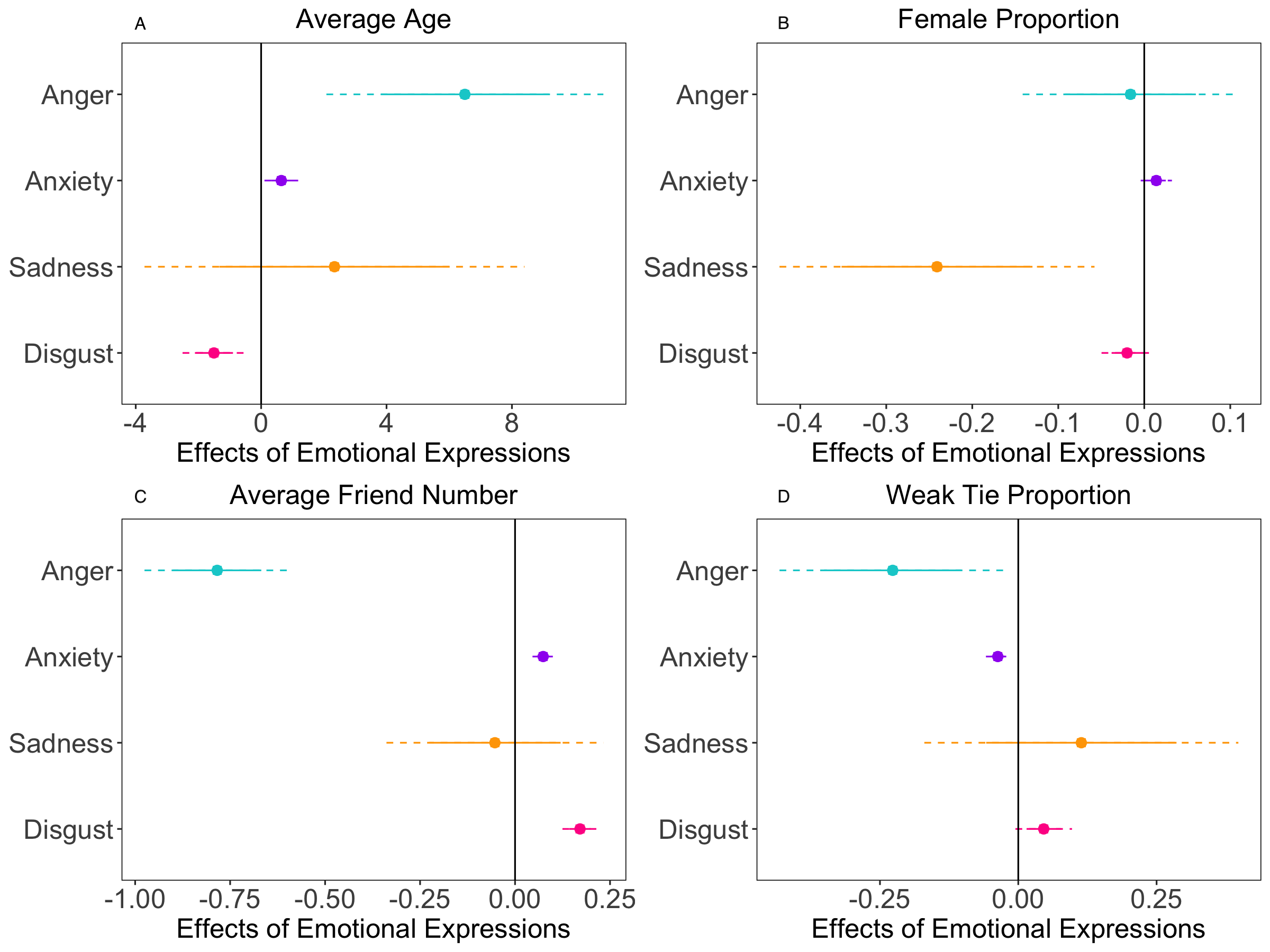}
\end{center}
\caption{The relationships between negative emotional expressions and cascade demographics and social ties: (A) average age of users, (B) female proportion of users, (C) average friends number of users, and (D) weak tie proportion of users in the cascades.}
\scriptsize Note: The $x$ axis displays the size of model coefficients. The $y$ axis displays four negative emotional expressions. The squares in the middle of the intervals represent mean values. The intervals indicated by dashed lines represent 90\% confidence intervals. The intervals indicated by solid lines represent mean values plus or minus 1 unit of their standard deviations.
\label{fig:emo_med}
\end{figure*}

We further investigated the effects of emotional expressions embedded in articles on the demographic characteristics (i.e., average age, proportion of females, and network degree) of all users involved in the cascades and the social ties (i.e., proportion of weak ties) among them, controlling for the effects of positive emotional expressions (See Appendix \ref{appendix: emopos_demo} for details).
% One may hypothesize that the characteristics of the individuals involved in a cascade, weak ties or the mechanism of social reinforcement may explain why articles with a higher level of anxiety, love, and sadness diffuse significantly differently in the online social network. To test this hypothesis, we perform mediation analysis  \citep{baron1986moderator}, by analyzing the demographic and network characteristics of more than 6 million users involved in the cascades and the social ties between them. 
% We find significant mediating effects of the average age, average network centrality (measured by users' average friend number), and social relations (measured by the proportion of weak ties) among the users in a cascade, respectively, on the relationships between each discrete emotion and each cascade dimension. The correlations among these three mediators are relatively small (less than 0.13), which guarantees the independence of these mediators.
See Table \ref{tab:emo_casprocess} and Figure \ref{fig:emo_med} for the results. 

The results show that all four negative emotional expressions play significant roles in affecting cascades' demographics and social ties. Previous literature suggests that age differences can result in different attitudes toward different emotions. For example, younger individuals can be more susceptible to negative low-arousal emotions due to gloating, such as disgust \citep{heath2001emotional}. At the same time, older users are more vulnerable to negative high-arousal emotions, such as anger and anxiety \citep{kensinger2008age}. Thus, the emotional expressions embedded in the articles may affect the age of the users involved in the cascades. Through the analysis, we find that articles with a higher degree of anger and anxiety expression embedded in content exhibit cascades that involve significantly older populations ($\beta_{anger}^{Average Age}=6.499$, $p<0.01$; $\beta_{anxiety}^{Average Age}=0.644$, $p<0.01$). In contrast, articles with a higher degree of expressions of disgust lead to a significantly younger population to participate in the cascades ($\beta_{disgust}^{Average Age}=-1.505$, $p<0.01$). 
% Our results suggest that articles embedded with a higher level of anxiety may be shared more in all four dimensions because it is better perceived by older users. And since older users may be more influential in affecting others' decision in participating in cascades, other users may be more willing to spread the associated articles. On the contrary, disgust is better perceived by younger users, which may lead to its insignificant effects on cascades dimensions. 
Previous research indicates that women are more likely to share positive information than are men \citep{lin2020examining}, and, therefore, women may be less engaged in cascades of articles with a high level of sadness. Our results consistently show that the degree of sadness expression has a significant negative impact on the proportion of females of the populations associated with the cascades ($\beta_{sadness}^{Female Proportion}=-0.241$, $p<0.01$). 
%Since males may be comparably less willing to share information, articles associated with a higher level of sadness may be less transmitted.  

We then explored how emotional expressions affect the network degree of individuals involved in cascades, measured by the average number of friends in the observation period \citep{jackson2008}. Empirical evidence suggests that users with a higher network degree enjoy higher social status and social influence. They tend to share content with more anxiety and less anger to maintain their status and influence, as expressions of anxiety are associated with an image of high cognitive capability and prudence \citep{yin2014anxious}. In contrast, expressions of angry correlate with a lack of self-control, low cognitive capability, and irrationality \citep{xiao2018social}. Further, people tend to consume stories about others as related to experiences of disgust based on a need for social comparison and to gloat \citep{heath2001emotional}. High-network-degree users, who tend to be more sociable, may share such stories to entertain their audience, establish an image of being humorous, and maintain social connections. Our results suggest that both the degree of expressions of anxiety and disgust positively affect the average number of friends of the users who participate in the cascades ($\beta_{anxiety}^{Average Friend Number}=0.074$, $p<0.001$; $\beta_{disgust}^{Average Friend Number}=0.171$, $p<0.001$). A higher degree of expressions of anger, however, leads to cascades with a population of fewer friends on average ($\beta_{anger}^{Average Friend Number}=-0.784$, $p<0.001$). 
%Our results suggest that articles embedded with more anxiety may be spread more because they are more shared by users with greater network centrality, who are more influential in broadcasting information. Meanwhile, articles with a higher level of anger are less shared in that users with more friends seldom share these articles.  

Finally, we investigated whether articles with different embedded emotions spread through different social ties (i.e., strong or weak ties). People are less motivated toward impression management with their strong-tie friends, and, thus, they may feel more comfortable with sharing content with anger with their close contacts. In addition, content with expressions of anxiety is perceived as more urgent and useful \citep{yin2014anxious}. Therefore, people may share such content with more important contacts (i.e., strong-tie friends).  Our results show that articles with a higher degree of anger and anxiety expressions will result in a lower proportion of weak ties among users who participate in the cascades ($\beta_{anger}^{Weak Tie Proportion}=-0.227$, $p<0.05$; $\beta_{anxiety}^{Weak Tie Proportion}=-0.037$, $p<0.001$). In summary, these results are consistent with our theoretical predictions developed in Section \ref{lr:cv}.

% Similar to our analysis of the effects of negative emotional expressions on structural properties of cascades, we also controlled for the positive emotional expressions embedded in the articles (please see Appendix \ref{appendix: emopos_demo} for details).
%Our results imply that articles with a higher level of anger are transmitted more through strong ties, which may explain the negative effects of anger on cascade size and dimension to some extent.
%%%%%%%%%%%%%%%%%%%%%%%%%%%%%%%%%%%%%%%%%%%%%%%%%%%%%%%%%%%%%%%%%%%%%%%%%%%%%%%%%%%%%%
\begin{table}
\centering
  \caption{\footnotesize Effects of Emotional Expressions on Demographics and Social Ties of Cascades}
     \label{tab:emo_casprocess}
 \begin{adjustbox}{width={0.8\textwidth},totalheight={1\textheight},keepaspectratio}
 \begin{threeparttable}
  \begin{tabular}{l*{5}{D{.}{.}{-1}}}
  %\begin{tabularx}{\textwidth}{@{} l
  %*{5}{S}}
    \toprule
    \multicolumn{1}{c}{\textbf{ } }    & \multicolumn{1}{c}{\textbf{Average Age}}     & \multicolumn{1}{c}{\textbf{Female Proportion}} & \multicolumn{1}{c}{\textbf{Average Friend}} & \multicolumn{1}{c}{\textbf{Weak Tie Proportion}} \\
    \midrule
    \multicolumn{1}{l}{\textbf{Variable}}\\
    \cmidrule(r){1-5}
    Anger&6.499**&	-0.016&	-0.784***&	-0.227*	    \\
         & (2.677)&	(0.076)&	(0.116)&	(0.124)\\
    Anxiety&0.644**&	0.014&	0.074***&	-0.037***\\
           &(0.326)&	(0.011)&	(0.017)&	(0.013) \\
    Sadness&2.34&	-0.241**&	-0.053&	0.114\\
           &(3.673)&	(0.111)&	(0.173)&	(0.172) \\
    Disgust&-1.505**&	-0.02&	0.171***&	0.046 	   \\
           &(0.607)&	(0.018)&	(0.028)&	(0.031) \\
    % Joy&-0.004&     -0.003* &  -0.002 &  -0.002\\
    %   &(0.003) &   (0.002)&  (0.002)&  (0.002)\\
    % Love&0.006*** &   0.004*** & 0.002* & 0.004***\\
    %     &(0.002)  &  (0.001) & (0.001)&  (0.001) \\
    % Surprise&-0.009  &    0.009  & 0.017** &  -0.004 \\
    %         &(0.011)  &  (0.008)&  (0.008) & (0.007) \\
    % Anticipation&0.004 &     -0.003 &  -0.004 &  -0.001 \\
    %             & (0.005) &   (0.003) & (0.002)&  (0.003) \\
    \cmidrule(r){1-5}
    % \multicolumn{1}{l}{\textbf{Article-level Controls\tnote{3}}}\\
    % \cmidrule(r){1-6}
    % %Originality&0.093***&0.06***&0.051***&-0.052&0.056***\\
    % %Info. Uniqueness&-0.057***&-0.02***&0.021***&0.079&-0.032***\\
    % Article Length&0.008&0.034***&0.041***&-0.294***&-0.004\\
    % \# of Images&0.09***&0.042***&0.035***&-0.095*&0.05***\\
    % \# of Videos&0.163***&0.103***&0.096***&0.296***&0.081***\\
    % Post at Weekends&-0.004&0.006***&0.008***&0.064*&-0.001\\
    % \# of Comments&0.218***&0.114***&0.099***&-0.003&0.101***\\
    % Reward Function&0.03***&0.029***&0.028***&-0.106*&0.02***\\					
    % \cmidrule(r){1-6}
    % \multicolumn{1}{l}{\textbf{Publisher-level Controls}}\\
    % \cmidrule(r){1-6}
    % Ave. \# of Posts&-0.195***&0.035***&0.109***&-0.597***&-0.084***\\
    % \# of Followers&0.35***&0.575***&0.643***&1.484***&0.132***\\
    % Type: Individual&-0.009&0.016*&0.022**&0.252**&-0.005\\
    % Type: Business&-0.123***&-0.087***&-0.079***&0.348*&-0.089***\\
    % Type: Gov.&-0.066***&-0.022&-0.008&0.704***&-0.05***\\
    % Type: Media&-0.096*&-0.075&-0.055&0.449&-0.066**\\
    % \cmidrule(r){1-6}
    %\hline
    \textbf{Positive emotional expressions}& \checkmark & \checkmark & \checkmark & \checkmark\cr
    \textbf{Article-level characteristics}& \checkmark & \checkmark & \checkmark & \checkmark\cr
    \textbf{Publisher fixed effects}& \checkmark & \checkmark & \checkmark & \checkmark\cr
    \bottomrule
  \end{tabular}
%   \begin{tablenotes}
%   %\small
%   \scriptsize
%     \item[1] \textit{Significance level: * $p < 0.05$, ** $p<0.01$, *** $p < 0.001$}
%     % \item[2] \textit{Given the right-skewed distributions of size and maximum breadth, we did a log-transformation on these two variables.}
%     \item[2] \textit{Positive emotional expressions, i.e., joy, love, surprise, and anticipation, are controlled.}
%     \item[3] \textit{All of the article-level variables(i.e. article length, number of images and videos embedded in the article, whether the articles was posted during a weekend, number of comments and topic distribution of the article) are controlled.}
%     % \item[5] \textit{SV indicates structural virality.}
%     \item[4] \textit{Standard error of estimation.}
%     \end{tablenotes}
\end{threeparttable}
\end{adjustbox}
\end{table}
%%%%%%%%%%%%%%%%%%%%%%%%%%%%%%%%%%%%%%%%%%%%%%%%%%%%%%%%%%%%%%%%%%%%%%%%%%%%%%%%%%%%%%

\section{Robustness Checks}
A battery of tests validated our results and confirm their robustness. 
%\subsection{Emotion Received by Readers }
First, although emotions were expressed and detected in the content of articles, it was still uncertain whether and how readers received these emotions. We therefore analyzed the emotions expressed in the articles' comments to examine how readers receive the expressed emotions. We first selected out the articles that expressed an extreme emotion. In particular, for each of the four negative emotional expressions, we selected articles with only one of the emotional expression intensity scores as higher than 1.96 and others as lower than 1.96. Our results demonstrate that the audience can receive the same emotion expressed in the article, which is consistent with the prior research on emotional contagion \citep[e.g.,][]{hatfield1993emotional,barsade2002ripple}\footnote{Because all emotional expression intensity scores were standardized as z-scores, the threshold of 1.96 indicates that the articles selected out have only one of the emotion intensity scores that is statistically higher than that of an an average article at a 95\% significance level, whereas other emotion intensity scores are not statistically significantly higher than that of an average article. Our results show that the most expressed emotion in comments is consistent with the emotion most expressed in the associated article (See Figure \ref{fig:CommentNew} in Appendix \ref{appendix: Robust Figures}). For example, if anxiety is the most expressed emotion in an article, anxiety would be, on average, the most expressed emotion in the comments.}. Our results also show that the audience may have other emotions different from the emotion most expressed in the content. This suggests that emotional contagion is not the only mechanism that drives an audience's emotions, which is consistent with the EASI framework \citep{van2009emotions}. Second, our sample includes articles with only a few words but with videos. The emotions in these articles may be expressed mainly in videos instead of the text, although, in most cases, the emotions expressed are consistent between the text and video. Our emotion analysis approach focused exclusively on emotional expressions embedded in text; thus, we dropped those articles with videos and that were shorter than 90\% of the articles in length, and replicated our analysis with this new sample (see Table \ref{tab:novideo} in Appendix \ref{appendix: Robust Tables}). The results are qualitatively consistent with our main results. Third, information novelty is arguably an influential factor in information diffusion \citep{berger2012makes,vosoughi2018spread}, and novel information may appear with emotional expressions (e.g., surprise, disgust) \citep{vosoughi2018spread}. We add \textit{article originality} (a verification by the WeChat platform that indicated officially whether an article is original) to our main model to control for information novelty. We use a dummy variable for which an article is coded as 1 if it is original and as 0 if it is reproduced, verified by the WeChat platform. The results are shown in Table \ref{tab:origin} in Appendix \ref{appendix: Robust Tables}, and the effects are qualitatively consistent with the previous results after controlling for article originality. Finally, we implemented a placebo test through random treatment analysis to further address the possible false significance concern. Specifically, we shuffled the emotion vectors and randomly assigned them to the articles in our sample. We then estimated our partial-linear instrumental variable model with the new dataset. We repeated this procedure ten times and all results show that the effects of negative emotional expressions are insignificant, which gives us confidence to argue that our estimated significant effects are not driven by large sample size\footnote{Unfortunately, we were unable to get the distribution in our limited amount of time due to time complexity and large sample size. The time complexity of XGBoost algorithm that we used in the double machine learning framework is $O(Kdx_0\log{ n})$ (here  $x_0$ denotes the number of non-missing entries in the training data, $d$ denotes the maximum depth of the tree and $K$ denotes the total number of trees). Empirically it took us over one week per round, and, therefore, we were unable to repeat it thousands of times to get the distribution.}.

\section{Discussion}

Our work shows that negative emotions, especially anxiety, anger, and sadness, embedded in online content can lead to differential diffusion patterns (i.e., cascade size, depth, maximum breadth, and structural virality). We are among the first to investigate the effects of negative emotional expressions on the characteristics of individuals (i.e., age, gender, and network degree) and social ties (i.e., strong and weak) involved in the cascading process. Our results suggest that expressing more negative emotions in online content can lead to differing compositions of individuals and social ties in the cascades. 
% For example, the articles with more expressions of anxiety tend to spread among individuals of a greater age and network degree (more friends in their local social networks) and among strong ties. The articles with more expressions of anger are diffused more through older individuals with fewer friends and weak ties. Females tend to share less information that expresses more sadness. Young people like to transmit the content that expresses disgust. 
Our novel findings not only shed light on content marketing and regulation, but also add to the literature in social psychology and social networks. 
% Our results also provide opportunities for future research to understand the social-psychological mechanisms of information diffusion.

% More importantly, we contribute to the literature of emotions in online content diffusion \citep[e.g.,][]{heath2001emotional,berger2012makes,brady2017emotion} from the following aspects. First, the existing studies on how emotions shape the diffusion of a few specific types of online content, i.e., political, moralized, and cancer tweets \citep{stieglitz2013emotions,brady2017emotion,wang2020negative} and New York Times articles shared through email \citep{berger2012makes}. We study 387,486 articles randomly sampled from a massive-scale online social network, covering a variety of topics and including both short texts and long articles, which deepens our understanding of how emotions shape general online content diffusion.

Methodologically, we \textit{causally} identified the impact of negative emotional expressions on online content diffusion with large-scale observational data, differing from past studies built on predictive models \citep{berger2012makes,stieglitz2013emotions,brady2017emotion,wang2020negative} or small-scale lab experiments \citep{heath2001emotional,berger2012makes}.
Our approach addresses the endogeneity issues caused by unobservable factors through instrumental variables, and also flexibly controls for the potentially non-linear effects of observable factors using machine learning techniques. Our results are of higher generalizability than those of previous studies, such as lab experiments, due to our randomly selected large-scale and comprehensive sample. We implemented a state-of-the-art, scalable, and domain-adaptive approach to accurately detect emotional expressions in massive online content. We also are one of the first to precisely measure cascade structures, as well as demographics and social relationships involved in the cascades. 

% advance the identification of the critical role of negative emotional expressions in online content diffusion, from both the causal inference and measurement perspectives. From the perspective of causal inference, differing from past studies that are based on predictive models \citep{berger2012makes,stieglitz2013emotions,brady2017emotion,wang2020negative} or small-scale lab experiments \citep{heath2001emotional,berger2012makes}, we present the first results of causal inference based on large-scale empirical data.

% Our approach flexibly controls for the potentially non-linear effects of observable factors using machine learning techniques and mitigates the endogeneity issues caused by unobservable factors. Thanks to the randomness, comprehensiveness and scale of our data, we believe that our results are of higher external validity than those of lab experiments. From the perspective of measurement, we use a state-of-the-art, scalable and domain-adaptive approach to facilitate more accurate detection of emotional expressions in massive content. This paper is also the first in this literature to measure cascade structures, and demographics and social relationships involved in the cascades. Understanding the relationship between emotional expressions and a variety of cascade characteristics is informative for content marketing and regulation. The empirical results on content with emotions spreading through heterogeneous individuals and social ties further open new opportunities of understanding the underlying sociopsychological mechanisms for future research.

Theoretically, we proposed a customized EASI framework \citep{van2009emotions} to theorize the impact of negative emotional expressions on the social transmission of online content. The dominant theories in the literature, such as arousal driving social transmission \citep{berger2012makes}\footnote{This theory argues that high-arousal negative emotional expressions (anger and anxiety) cause online content to go viral \citep{berger2012makes}. We find that although the effects of anxiety on diffusion are positive (consistent with the prediction of \cite{berger2012makes}), expressions of anger reduce virality.}  and negativity bias theory \citep{stieglitz2013emotions,hennig2015does}, cannot sufficiently explain the heterogeneous impacts of negative emotional expressions on information diffusion.
Our customized EASI framework incorporates the theories of emotional contagion \citep[e.g.,][]{hatfield1993emotional,barsade2002ripple,del2016echo}, emotion-induced approach-avoidance motivation \citep{elliot2013approach}, cognitive appraisal \citep{lerner2015emotion}, and social comparison \citep{heath2001emotional}, some of the most important theories of emotional expressions in the social psychology literature, and links them to the information diffusion process and outcomes. 
We highlighted two key processes that drive the effects of emotional expressions on content diffusion, i.e., cognitive inferences and affective reactions. This framework enables us to deeply interpret the impacts of negative emotional expressions. 

In addition, our work provides valuable insight for practitioners such as content producers, platform managers and marketers. First, content producers can rely on the expression of negative emotions embedded in the content to promote or inhibit the social dissemination of online information, such as news articles, product advertisements, and campaign activities. For example, they can expand the spread of their advertisements on social media networks by adding expressions of anxiety. In contrast, sadness and anger should not be overexpressed in the content to be propagated in social networks. 
Second,  to reduce negative emotional content, platform managers should pay special attention to anxiety than to other negative emotions. Content with anxiety is shown to spread more broadly and more virally, indicating that the online community (especially a community with closely connected older users) is more vulnerable and reactive to anxious content. Anecdotally, we observe that social media is flooded with more information that expresses anxiety, such as the content about new virus mutations and pandemic developments.  

Third, our results shed light on the strategies for marketers to promote online information with different expressions of negative emotions in social networks. Marketers are not necessarily with many followers on social media.\footnote{Even if marketers have their own social media accounts (e.g., official accounts), they may not have enough followers. Instead, marketers often have to rely on and pay for popular publishers or key opinion leaders (KOLs) to promote content. Publishers or KOLs then propagate the marketer-generated content to (all or some of) their followers, spreading the content on social networks. This is, in fact, one of the most important business models for publishers or KOLs to generate revenue on social media platforms, such as WeChat.} To spread their content, marketers often need to choose the right publishers or key opinion leaders (KOLs), and then apply broadcasting and viral strategies to promote content (e.g., advertisements, elections) after getting the initial user pool from publishers or KOLs. 
Specifically, our results can provide guidance for matching the content of negative emotional expressions with the users' characteristics (individual characteristics and social ties) to increase the content diffusion in social media networks in at least two aspects. First, we can help select appropriate seeds, such as the characteristics of publishers and KOLs that provide the initial user pool to spread the content, and their particular followers to transmit the content of negative emotional expressions socially in networks. Second, our findings also suggest the appropriate strategies, such as broadcasting and viral strategies, to be applied among the targeted users (e.g., followers) to maximize the diffusion. For example, expressions of anxiety can be leveraged by content providers, along with targeting older users with more friends and, especially, close friends (strong ties) as seeds to promote the information in social networks. The spread of angry content can be strengthened, if it seeds among strong ties and older people with fewer friends. Further, broadcasting content with users of a high network degree can be used to spread online content that expresses more anxiety and disgust. A viral strategy (e.g., social referrals) can facilitate the spread of the content expressing more anxiety, as the articles with greater expressions of anxiety lead to greater depth and structural virality of cascades.

Our study is not without limitations. First, existing research in psychology has not achieved agreement on the basic discrete emotion categories for human beings \citep{plutchik1980emotion,lerner2015emotion}. Although we focused on four negative discrete emotions that are well defined in the literature and commonly expressed in online content, future research could investigate whether other discrete emotions, for example, contempt, fear, shame, and guilt, can affect the information-cascading process and how they interact with social or psychological processes. 
%Second, future work may investigate the psychological processes involved in each relationship that we find. For example, it would be valuable to answer the question regarding why content with certain emotions tends to spread through various individuals and social ties.
Second, our models follow the research \citep{berger2012makes,nguyen2020influence} that assumes the effects of discrete emotions are additive in a linear way. Future research can use an experimental approach to manipulate different portfolios of discrete emotions and test whether the interaction of discrete emotions has an impact on content diffusion. 
Third, we did not explore the nested effects of publishers within different periods of time because our research utilizes a standard two-way fixed effects model to better account for endogeneity concerns when discovering the causal relationships. The publisher-time nested effects are not identifiable in our fixed effects models. Future research can examine how different publishers within different periods of time influence diffusion.
Fourth, future research could examine how the effects of emotions in content diffusion vary across different topics. Understanding this will provide content producers with a means to write and promote their work in social media in specific topics.\footnote{We did not explore the interaction effects of emotions and topics in article diffusion, as, in our context, articles with several topic tags do not have a single clear classification.} 
Fifth, we chose not to use comment data in our model because it has limitations and cannot contribute to a clean causal identification.\footnote{To begin with, comments are generated sequentially. The comments generated later can be affected by the ones generated earlier. Comments generated later also may be endogenous to content diffusion outcomes. For example, people may provide more comments on widely spread content. Moreover, in our context, although we can observe all comments from the company's database, not all comments are visible to readers. The writer usually selects only a few comments and presents them to readers. Such a self-selection mechanism makes it challenging to identify how comments affect content diffusion.} Future research can examine the roles that the comments play in content diffusion. Sixth, we chose to focus on the impact of \textit{emotional expression} embedded in content, rather than the emotions experienced by the readers, on content diffusion. This is because, in our context, emotional expression can be measured empirically and the understanding the impact of emotional expression is actionable for writers who can decide on the degree of emotions expressed in their content, whereas emotions experienced by the readers are unobservable and are difficult to predict by writers. 
Future research can explore how the emotional expressions embedded in online content affect readers' emotional states. Finally, state-of-the-art emotion analysis approach in empirical research is from the word level, and our study relies on this technology. Also, it is less ambiguous for writers to take actions based on the findings of the impact of word-level emotionality rather than article-level emotionality. Nevertheless, large-scale article-level emotion analysis is an important question for future natural language process research to address.
% Finally, considering the managerial importance of negative emotions, we choose to focus mainly on negative discrete emotional expressions on content diffusion, with controlling the effects of positive discrete emotional expressions (the results of which are reported in Appendix \ref{appendix: emopos_results}). Future research can further investigate the underlying mechanisms and theories for positive emotions in online content diffusion.

% This work also generates a number of interesting questions for future research. For example, how to understand the effects of emotions in each dimension of the information cascades from a theoretical perspective? Why are certain emotions linked to some of the dimensions, whereas other emotions are not? To answer these questions requires joint efforts from both psychology and computational social science research.

\bibliographystyle{informs2014} % outcomment this and next line in Case 1
%\bibliography{msref}
%\bibliography{msref,SIbib} % if more than one, comma separated
\bibliography{Emotion_ContentDiffusion}

\begin{thebibliography}{65}
\providecommand{\natexlab}[1]{#1}
\providecommand{\url}[1]{\texttt{#1}}
\providecommand{\urlprefix}{URL }

\bibitem[{Adams et~al.(2006)Adams, Ambady, Macrae, \protect\BIBand{}
  Kleck}]{adams2006emotional}
Adams RB, Ambady N, Macrae CN, Kleck RE (2006) Emotional expressions forecast
  approach-avoidance behavior. \emph{Motivation and emotion} 30(2):177--186.

\bibitem[{Bailis \protect\BIBand{} MacCoun(1996)}]{bailis1996estimating}
Bailis DS, MacCoun RJ (1996) Estimating liability risks with the media as your
  guide: A content analysis of media coverage of tort litigation. \emph{Law and
  Human Behavior} 20(4):419--429.

\bibitem[{Banerjee et~al.(2013)Banerjee, Chandrasekhar, Duflo,
  \protect\BIBand{} Jackson}]{banerjee2013diffusion}
Banerjee A, Chandrasekhar AG, Duflo E, Jackson MO (2013) The diffusion of
  microfinance. \emph{Science} 341(6144):1236498.

\bibitem[{Barsade(2002)}]{barsade2002ripple}
Barsade SG (2002) The ripple effect: Emotional contagion and its influence on
  group behavior. \emph{Administrative science quarterly} 47(4):644--675.

\bibitem[{Berger \protect\BIBand{} Milkman(2012)}]{berger2012makes}
Berger J, Milkman KL (2012) What makes online content viral? \emph{Journal of
  marketing research} 49(2):192--205.

\bibitem[{Blei et~al.(2003)Blei, Ng, \protect\BIBand{} Jordan}]{blei2003latent}
Blei DM, Ng AY, Jordan MI (2003) Latent dirichlet allocation. \emph{Journal of
  machine Learning research} 3(Jan):993--1022.

\bibitem[{Bollen et~al.(2011)Bollen, Mao, \protect\BIBand{}
  Zeng}]{bollen2011twitter}
Bollen J, Mao H, Zeng X (2011) Twitter mood predicts the stock market.
  \emph{Journal of computational science} 2(1):1--8.

\bibitem[{Brady et~al.(2019)Brady, Wills, Burkart, Jost, \protect\BIBand{}
  Van~Bavel}]{brady2019ideological}
Brady WJ, Wills JA, Burkart D, Jost JT, Van~Bavel JJ (2019) An ideological
  asymmetry in the diffusion of moralized content on social media among
  political leaders. \emph{Journal of Experimental Psychology: General}
  148(10):1802.

\bibitem[{Brady et~al.(2017)Brady, Wills, Jost, Tucker, \protect\BIBand{}
  Van~Bavel}]{brady2017emotion}
Brady WJ, Wills JA, Jost JT, Tucker JA, Van~Bavel JJ (2017) Emotion shapes the
  diffusion of moralized content in social networks. \emph{Proceedings of the
  National Academy of Sciences} 114(28):7313--7318.

\bibitem[{Chen \protect\BIBand{} Guestrin(2016)}]{chen2016xgboost}
Chen T, Guestrin C (2016) Xgboost: A scalable tree boosting system.
  \emph{Proceedings of the 22nd acm sigkdd international conference on
  knowledge discovery and data mining}, 785--794.

\bibitem[{Chernozhukov et~al.(2018)Chernozhukov, Chetverikov, Demirer, Duflo,
  Hansen, Newey, \protect\BIBand{} Robins}]{chernozhukov2018double}
Chernozhukov V, Chetverikov D, Demirer M, Duflo E, Hansen C, Newey W, Robins J
  (2018) Double/debiased machine learning for treatment and structural
  parameters.

\bibitem[{Clark \protect\BIBand{} Taraban(1991)}]{clark1991reactions}
Clark MS, Taraban C (1991) Reactions to and willingness to express emotion in
  communal and exchange relationships. \emph{Journal of Experimental Social
  Psychology} 27(4):324--336.

\bibitem[{Coviello et~al.(2014)Coviello, Sohn, Kramer, Marlow, Franceschetti,
  Christakis, \protect\BIBand{} Fowler}]{coviello2014detecting}
Coviello L, Sohn Y, Kramer AD, Marlow C, Franceschetti M, Christakis NA, Fowler
  JH (2014) Detecting emotional contagion in massive social networks.
  \emph{PloS one} 9(3):e90315.

\bibitem[{Del~Vicario et~al.(2016)Del~Vicario, Vivaldo, Bessi, Zollo, Scala,
  Caldarelli, \protect\BIBand{} Quattrociocchi}]{del2016echo}
Del~Vicario M, Vivaldo G, Bessi A, Zollo F, Scala A, Caldarelli G,
  Quattrociocchi W (2016) Echo chambers: Emotional contagion and group
  polarization on facebook. \emph{Scientific reports} 6(1):1--12.

\bibitem[{Dev et~al.(2019)Dev, Karahalios, \protect\BIBand{}
  Sundaram}]{dev2019quantifying}
Dev H, Karahalios K, Sundaram H (2019) Quantifying voter biases in online
  platforms: An instrumental variable approach. \emph{Proceedings of the ACM on
  Human-Computer Interaction} 3(CSCW):1--27.

\bibitem[{Dolan(2002)}]{dolan2002emotion}
Dolan RJ (2002) Emotion, cognition, and behavior. \emph{science}
  298(5596):1191--1194.

\bibitem[{Edelman et~al.(1992)Edelman, Abraham, \protect\BIBand{}
  Erlanger}]{edelman1992professional}
Edelman LB, Abraham SE, Erlanger HS (1992) Professional construction of law:
  The inflated threat of wrongful discharge. \emph{Law and Society Review}
  47--83.

\bibitem[{Elliot et~al.(2013)Elliot, Eder, \protect\BIBand{}
  Harmon-Jones}]{elliot2013approach}
Elliot AJ, Eder AB, Harmon-Jones E (2013) Approach--avoidance motivation and
  emotion: Convergence and divergence. \emph{Emotion Review} 5(3):308--311.

\bibitem[{Finkenauer(1998)}]{finkenauer1998secrets}
Finkenauer C (1998) Secrets: Types, determinants, functions, and consequences.
  \emph{Unpublished doctoral dissertation, University of Louvain at
  Louvain-la-Neuve, Louvain-la-Neuve, Belgium} .

\bibitem[{Goel et~al.(2015)Goel, Anderson, Hofman, \protect\BIBand{}
  Watts}]{goel2015structural}
Goel S, Anderson A, Hofman J, Watts DJ (2015) The structural virality of online
  diffusion. \emph{Manage Sci} 62(1):180--196.

\bibitem[{Granovetter(1977)}]{granovetter1977strength}
Granovetter MS (1977) The strength of weak ties. \emph{Social networks},
  347--367 (Elsevier).

\bibitem[{Hatfield et~al.(1993)Hatfield, Cacioppo, \protect\BIBand{}
  Rapson}]{hatfield1993emotional}
Hatfield E, Cacioppo JT, Rapson RL (1993) Emotional contagion. \emph{Current
  directions in psychological science} 2(3):96--100.

\bibitem[{Hausman \protect\BIBand{} Bresnahan(2008)}]{hausman20085}
Hausman JA, Bresnahan TF (2008) \emph{5. Valuation of New Goods under Perfect
  and Imperfect Competition} (University of Chicago Press).

\bibitem[{Heath et~al.(2001)Heath, Bell, \protect\BIBand{}
  Sternberg}]{heath2001emotional}
Heath C, Bell C, Sternberg E (2001) Emotional selection in memes: the case of
  urban legends. \emph{Journal of personality and social psychology}
  81(6):1028.

\bibitem[{Hennig-Thurau et~al.(2015)Hennig-Thurau, Wiertz, \protect\BIBand{}
  Feldhaus}]{hennig2015does}
Hennig-Thurau T, Wiertz C, Feldhaus F (2015) Does twitter matter? the impact of
  microblogging word of mouth on consumers’ adoption of new movies.
  \emph{Journal of the Academy of Marketing Science} 43(3):375--394.

\bibitem[{Jackson(2008)}]{jackson2008}
Jackson MO (2008) \emph{{Social and economic networks}}, volume~3 (Princeton
  University Press).

\bibitem[{Jiang et~al.(2018)Jiang, Drolet, \protect\BIBand{}
  Kim}]{jiang2018age}
Jiang L, Drolet A, Kim HS (2018) Age and social support seeking: Understanding
  the role of perceived social costs to others. \emph{Personality and Social
  Psychology Bulletin} 44(7):1104--1116.

\bibitem[{Keltner \protect\BIBand{} Haidt(1999)}]{keltner1999social}
Keltner D, Haidt J (1999) Social functions of emotions at four levels of
  analysis. \emph{Cognition \& Emotion} 13(5):505--521.

\bibitem[{Kensinger(2008)}]{kensinger2008age}
Kensinger EA (2008) Age differences in memory for arousing and nonarousing
  emotional words. \emph{The Journals of Gerontology Series B: Psychological
  Sciences and Social Sciences} 63(1):P13--P18.

\bibitem[{Kramer et~al.(2014)Kramer, Guillory, \protect\BIBand{}
  Hancock}]{kramer2014experimental}
Kramer AD, Guillory JE, Hancock JT (2014) Experimental evidence of
  massive-scale emotional contagion through social networks. \emph{Proceedings
  of the National Academy of Sciences} 111(24):8788--8790.

\bibitem[{Lerner et~al.(2015)Lerner, Li, Valdesolo, \protect\BIBand{}
  Kassam}]{lerner2015emotion}
Lerner JS, Li Y, Valdesolo P, Kassam KS (2015) Emotion and decision making.
  \emph{Annual review of psychology} 66:799--823.

\bibitem[{Lerner et~al.(2004)Lerner, Small, \protect\BIBand{}
  Loewenstein}]{lerner2004heart}
Lerner JS, Small DA, Loewenstein G (2004) Heart strings and purse strings:
  Carryover effects of emotions on economic decisions. \emph{Psychological
  science} 15(5):337--341.

\bibitem[{Lin \protect\BIBand{} Wang(2020)}]{lin2020examining}
Lin X, Wang X (2020) Examining gender differences in people’s
  information-sharing decisions on social networking sites. \emph{International
  Journal of Information Management} 50:45--56.

\bibitem[{Malik \protect\BIBand{} Hussain(2017)}]{malik2017helpfulness}
Malik M, Hussain A (2017) Helpfulness of product reviews as a function of
  discrete positive and negative emotions. \emph{Computers in Human Behavior}
  73:290--302.

\bibitem[{Mikolov et~al.(2013)Mikolov, Chen, Corrado, \protect\BIBand{}
  Dean}]{mikolov2013efficient}
Mikolov T, Chen K, Corrado G, Dean J (2013) Efficient estimation of word
  representations in vector space. \emph{arXiv preprint arXiv:1301.3781} .

\bibitem[{Nevo(2000)}]{nevo2000mergers}
Nevo A (2000) Mergers with differentiated products: The case of the
  ready-to-eat cereal industry. \emph{The RAND Journal of Economics} 395--421.

\bibitem[{Nguyen et~al.(2020)Nguyen, Calantone, \protect\BIBand{}
  Krishnan}]{nguyen2020influence}
Nguyen H, Calantone R, Krishnan R (2020) Influence of social media emotional
  word of mouth on institutional investors’ decisions and firm value.
  \emph{Manage Sci} 66(2):887--910.

\bibitem[{Plutchik(2001)}]{plutchik2001nature}
Plutchik R (2001) The nature of emotions: Human emotions have deep evolutionary
  roots, a fact that may explain their complexity and provide tools for
  clinical practice. \emph{American scientist} 89(4):344--350.

\bibitem[{Plutchik \protect\BIBand{} Kellerman(1980)}]{plutchik1980emotion}
Plutchik R, Kellerman H (1980) \emph{Emotion, theory, research, and experience}
  (Academic press).

\bibitem[{Quan \protect\BIBand{} Ren(2010)}]{quan2010blog}
Quan C, Ren F (2010) A blog emotion corpus for emotional expression analysis in
  chinese. \emph{Computer Speech \& Language} 24(4):726--749.

\bibitem[{Rim{\'e}(2009)}]{rime2009emotion}
Rim{\'e} B (2009) Emotion elicits the social sharing of emotion: Theory and
  empirical review. \emph{Emotion review} 1(1):60--85.

\bibitem[{Russell(1980)}]{russell1980circumplex}
Russell JA (1980) A circumplex model of affect. \emph{Journal of personality
  and social psychology} 39(6):1161.

\bibitem[{Saarim{\"a}ki et~al.(2016)Saarim{\"a}ki, Gotsopoulos,
  J{\"a}{\"a}skel{\"a}inen, Lampinen, Vuilleumier, Hari, Sams,
  \protect\BIBand{} Nummenmaa}]{saarimaki2016discrete}
Saarim{\"a}ki H, Gotsopoulos A, J{\"a}{\"a}skel{\"a}inen IP, Lampinen J,
  Vuilleumier P, Hari R, Sams M, Nummenmaa L (2016) Discrete neural signatures
  of basic emotions. \emph{Cerebral cortex} 26(6):2563--2573.

\bibitem[{Sargan(1958)}]{10.2307/1907619}
Sargan JD (1958) The estimation of economic relationships using instrumental
  variables. \emph{Econometrica} 26(3):393--415, ISSN 00129682, 14680262.

\bibitem[{Shi et~al.(2014)Shi, Rui, \protect\BIBand{}
  Whinston}]{shi2014content}
Shi Z, Rui H, Whinston AB (2014) Content sharing in a social broadcasting
  environment: evidence from twitter. \emph{MIS quarterly} 38(1):123--142.

\bibitem[{Song et~al.(2019)Song, Huang, Tan, \protect\BIBand{}
  Yu}]{song2019using}
Song T, Huang J, Tan Y, Yu Y (2019) Using user-and marketer-generated content
  for box office revenue prediction: Differences between microblogging and
  third-party platforms. \emph{Information Systems Research} 30(1):191--203.

\bibitem[{Song et~al.(2018)Song, Shi, Li, \protect\BIBand{}
  Zhang}]{song2018directional}
Song Y, Shi S, Li J, Zhang H (2018) Directional skip-gram: Explicitly
  distinguishing left and right context for word embeddings. \emph{Proceedings
  of the 2018 Conference of the North American Chapter of the Association for
  Computational Linguistics: Human Language Technologies, Volume 2 (Short
  Papers)}, 175--180.

\bibitem[{Stieglitz \protect\BIBand{} Dang-Xuan(2013)}]{stieglitz2013emotions}
Stieglitz S, Dang-Xuan L (2013) Emotions and information diffusion in social
  media—sentiment of microblogs and sharing behavior. \emph{Journal of
  management information systems} 29(4):217--248.

\bibitem[{Stock \protect\BIBand{} Yogo(2005)}]{16284}
Stock J, Yogo M (2005) \emph{Testing for Weak Instruments in Linear IV
  Regression}, 80--108 (New York: Cambridge University Press).

\bibitem[{Sun et~al.(2019)Sun, Qiu, Xu, \protect\BIBand{} Huang}]{sun2019fine}
Sun C, Qiu X, Xu Y, Huang X (2019) How to fine-tune bert for text
  classification? \emph{China National Conference on Chinese Computational
  Linguistics}, 194--206 (Springer).

\bibitem[{Tomkins(1962)}]{tomkins1962affect}
Tomkins SS (1962) \emph{Affect imagery consciousness: Volume I: The positive
  affects}, volume~1 (Springer publishing company).

\bibitem[{Van~den Bulte et~al.(2018)Van~den Bulte, Bayer, Skiera,
  \protect\BIBand{} Schmitt}]{van2018customer}
Van~den Bulte C, Bayer E, Skiera B, Schmitt P (2018) How customer referral
  programs turn social capital into economic capital. \emph{Journal of
  Marketing Research} 55(1):132--146.

\bibitem[{Van~Kleef(2009)}]{van2009emotions}
Van~Kleef GA (2009) How emotions regulate social life: The emotions as social
  information (easi) model. \emph{Current directions in psychological science}
  18(3):184--188.

\bibitem[{Van~Kleef(2010)}]{van2010emerging}
Van~Kleef GA (2010) The emerging view of emotion as social information.
  \emph{Social and Personality Psychology Compass} 4(5):331--343.

\bibitem[{Villas-Boas \protect\BIBand{} Winer(1999)}]{villas1999endogeneity}
Villas-Boas JM, Winer RS (1999) Endogeneity in brand choice models.
  \emph{Manage Sci} 45(10):1324--1338.

\bibitem[{Vosoughi et~al.(2018)Vosoughi, Roy, \protect\BIBand{}
  Aral}]{vosoughi2018spread}
Vosoughi S, Roy D, Aral S (2018) The spread of true and false news online.
  \emph{Science} 359(6380):1146--1151.

\bibitem[{Wang \protect\BIBand{} Lee(2020)}]{wang2020negative}
Wang X, Lee EW (2020) Negative emotions shape the diffusion of cancer tweets:
  toward an integrated social network--text analytics approach. \emph{Internet
  Research} .

\bibitem[{Wethington(2000)}]{wethington2000expecting}
Wethington E (2000) Expecting stress: Americans and the “midlife crisis”.
  \emph{Motivation and Emotion} 24(2):85--103.

\bibitem[{Wooldridge(2002)}]{Wooldridge2002}
Wooldridge JM (2002) {Econometric Analysis of Cross Section and Panel Data}.
  \emph{Booksgooglecom} 58(2):752, ISSN 09331719.

\bibitem[{Xiao et~al.(2018)Xiao, Zhang, \protect\BIBand{}
  Cervone}]{xiao2018social}
Xiao Y, Zhang H, Cervone D (2018) Social functions of anger: A competitive
  mediation model of new product reviews. \emph{Journal of Product Innovation
  Management} 35(3):367--388.

\bibitem[{Xue et~al.(2014)Xue, Fu, \protect\BIBand{} Shaobin}]{xue2014study}
Xue B, Fu C, Shaobin Z (2014) A study on sentiment computing and classification
  of sina weibo with word2vec. \emph{2014 IEEE International Congress on Big
  Data}, 358--363 (IEEE).

\bibitem[{Yang et~al.(2016)Yang, Zhang, Zhang, Mo, Li, Yu, \protect\BIBand{}
  Zhu}]{yang2016automatic}
Yang X, Zhang Z, Zhang Z, Mo Y, Li L, Yu L, Zhu P (2016) Automatic construction
  and global optimization of a multisentiment lexicon. \emph{Computational
  intelligence and neuroscience} 2016.

\bibitem[{Yin et~al.(2014)Yin, Bond, \protect\BIBand{} Zhang}]{yin2014anxious}
Yin D, Bond S, Zhang H (2014) Anxious or angry? effects of discrete emotions on
  the perceived helpfulness of online reviews. \emph{Mis Quarterly}
  38(2):539--560.

\bibitem[{Yin et~al.(2017)Yin, Bond, \protect\BIBand{} Zhang}]{yin2017keep}
Yin D, Bond S, Zhang H (2017) Keep your cool or let it out: Nonlinear effects
  of expressed arousal on perceptions of consumer reviews. \emph{Journal of
  Marketing Research} 54(3):447--463.

\bibitem[{Yu et~al.(2019)Yu, Yang, Huang, \protect\BIBand{}
  Tan}]{yu2019emotions}
Yu Y, Yang Y, Huang J, Tan Y (2019) Emotions in online reviews and product
  sales: Unifying empirical and theoretical perspectives. \emph{Available at
  SSRN 3497884} .

\end{thebibliography}

\clearpage
%% Here starts the e-companion (EC)
%%%%%%%%%%%%%%%%%%%%%%%%%%%%%%%%%%%%%%%%%%%%%%%%%%%%%%%%%%
\ECSwitch

%\ECDisclaimer
%%%%%%%%%%%%%%%%%%%%%%%%%%%%%%%%%%%%%%%%%%%%%%%%%%%%%%%%%%

%%% Main head for the e-companion
\ECHead{Online Appendix of Emotions in Online Content Diffusion}
% \begin{APPENDICES}
\setcounter{table}{0}
\counterwithin{table}{section}
\counterwithin{figure}{section}

\section{Construction of a Domain-Specific and Up-to-date Emotion Lexicon}
\label{appendix:construct}
Our approach requires an existing general emotion lexicon as a basic lexicon. Ren-CECps is an emotion lexicon based on 1,487 Chinese blog texts \citep{quan2010blog}. Each word $w_i$ in Ren-CECps has eight basic emotion types (i.e., surprise, joy, anticipation, love, anxiety, sadness, anger, and disgust). Each emotion class for each word is manually annotated from 0.0 to 1.0, which represents the emotion intensity expressed by the word. We denote the eight-dimension emotion-intensity vector as $v_i= \{e_j^i\}^{N=8}_{j=1}$. We select Ren-CECps for the following three reasons. First, as \cite{quan2010blog} noted, these eight types of emotions are most commonly expressed in Chinese blog texts; using these emotions decreases confusion in the emotion category selection. Second, the texts are annotated based on Chinese blogs. Blogs and WeChat articles are similar Chinese online content. Hence, we believe that Ren-CECps is more suitable for our context than are lexicons based on general Chinese texts (e.g., NTUSD, HowNet, DUT). Finally, Ren-CECps is manually annotated and statistically validated \citep{quan2010blog}.\footnote{Eleven annotators participated in the annotation work. According to \cite{quan2010blog}, the authors spent two months on the joint training of annotators and made annotation instructions. The authors also used a Kappa statistic to measure the pairwise agreement among 11 annotators. The Kappa coefficient of the agreement is a statistic adopted by the computational linguistics community as a standard measure for such a purpose. The agreement for emotion words is 0.785. Given the complexity of this annotation task, we believe that the annotations are reliable and valid.} Compared with automatically constructed emotion lexicons \citep[e.g.,][]{yang2016automatic}, the results of Ren-CECps are more precise and reliable. A total of 12,048 emotion words and their intensities were obtained from Ren-CECps as our basic emotion lexicon. 
    
Next, our approach requires word vectors that contain their semantic information. The word vectors can be derived by statistical language modeling (e.g. Word2Vec by \cite{mikolov2013efficient}). We used pre-trained word vectors by \cite{song2018directional}, who provide 200-dimension word vectors for over 8 million common Chinese words and phrases. These word vectors are pretrained on up-to-date, large-scale, and high-quality Chinese online content \citep{song2018directional}. Then, the similarity between the two words can be measured by the cosine similarity of the two corresponding word vectors \citep{mikolov2013efficient}. 

We follow the algorithm proposed by \cite{yu2019emotions} to extend the basic lexicon (denoted as $L_0$) to a domain-specific and up-to-date lexicon.\footnote{For more details, see Algorithms 1, 2, and 3 in \cite{yu2019emotions}.} First, we randomly split the basic lexicon $L_0 = \{(w_i, v_i)\}^N_{i=1}$ into training (90\%) and test (10\%) sets, i.e., $L_0^{tr}$ and $L_0^{te}$. Second, we construct $W$ as a set of all unique words in our sampled articles, except stopping words. From $W$, we construct a subset $P$ that contains potential emotion words. Specifically, for each word $w_i \in W$, if $w_i$ is not in $L_0^{tr}$, we get $M$ words that are semantically most similar to $w_i$: $\{w_{i,j}\}^M_{j=1}$, by using the pre-trained word vectors. $M$ is chosen as 100, as suggested by \cite{xue2014study} and \cite{yu2019emotions}. If there exists $j$ such that $w_{i,j} \in L_0^{tr}$, $w_{i,j}$ is considered a potential emotion word \citep{xue2014study}, and we add $w_{i,j}$ to $P$. Third, we iteratively extend the basic lexicon $L_0$. We define an algorithm $f(w_i;L, \theta)$ that maps a word $w_i$ to an emotion-intensity vector $v_i = (e_1, e_2, ..., e_8)$, where $e_i \in [0,1]$. As we detail later, $f$ is conditional on parameter $\theta$ and lexicon $L$. For each word $w_i \in P$, we use $f$ to map $w_i$ to an emotion-intensity vector $v_i$. If $v_i$ has at least one positive intensity value, we add a word-intensity pair ($w_i$, $v_i$) to the basic lexicon. After one iteration, we check whether the number of words in the basic lexicon
increases. If so, we repeat the iteration until the number of words converges.

Next, we introduce the algorithm $f$. The rationale of $f$ originates from the K-nearest neighbor algorithm, but we customize it in our task and introduce hyper-parameter $\alpha$ to control for noise. For a word $w$, we construct set $S_w$ that contains $K$-most similar words by using the pre-trained word vectors. We then check the intersection of $S_w$ and emotion lexicon $L$, denoted as $S_w^L$. If $S_w^L$ is non-empty, we use weighted average of the emotion intensities of emotion words in $S_w^L$ with respect to their similarity to $w$ to determine the emotion intensities of $w$. We compress emotion intensities that are lower than $\alpha$ to zero to reduce noise. The parameter $\theta = (K, \alpha)$ is chosen as (5, 0.15), which is the optimal parameter suggested by \cite{yu2019emotions}. 

We use out-of-sample testing to validate the approach. After each iteration $i$, we obtain an extended lexicon $L_i$. For each word $w$ in the test set $L_0^{te}$, we use $f(w; L_i, \theta)$ to generate a predicted emotion-intensity vector $v$ for $w$. Then, we use mean
absolute error (MAE) to evaluate the error between $v$ and the human-annotated value vector $v^0$ provided by \cite{quan2010blog}.

\begin{figure}
  \begin{center}
      \includegraphics[width=0.6\textwidth]{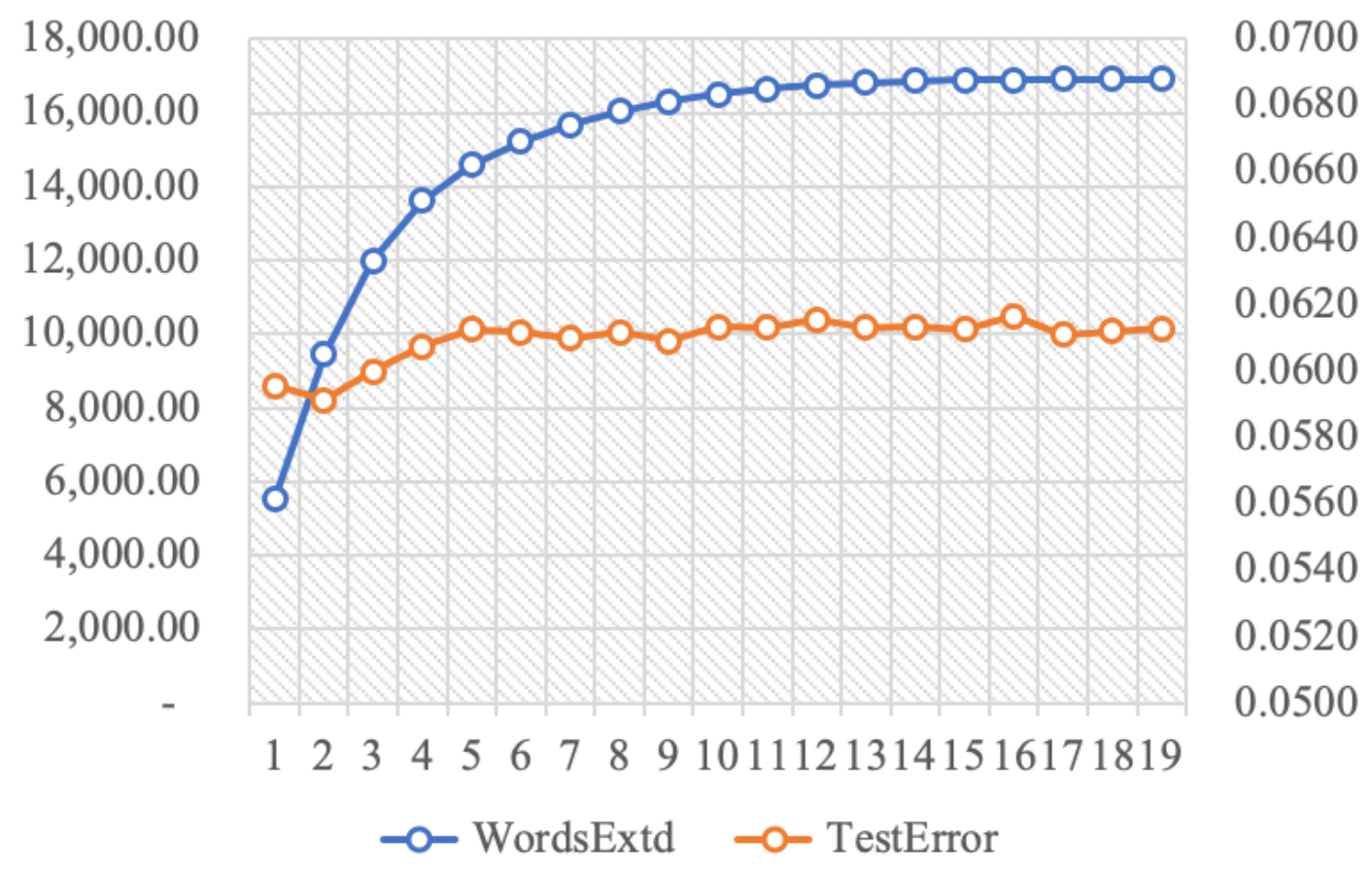}
  \end{center}
  
  \caption{Iterations, words extended and testing errors.}
  \scriptsize The figure illustrates the number of extended words and prediction errors for each iteration. When the number of iterations was 20, the number of emotion words converged. During the iteration, the prediction error varied slightly, from 0.0580 to 0.0620. The low and stable prediction error level ensured the validity of the entire iteration process and the accuracy of the lexicon. A total of 16,921 new words were found through this process.
  \label{fig:Iterations}
\end{figure}

Figure \ref{fig:Iterations} illustrates the number of extended words and the prediction error for each iteration. When the number of iterations is 20, the number of emotion words converges. During the iteration, the prediction error varied slightly, from 0.0580 to 0.0620. The low and stable prediction error level ensured the validity of the entire iteration process and the accuracy of the lexicon. A total of 16,921 new words were found after this process. This result confirms the necessity of constructing a domain-specific lexicon, without which 58.4\% of unique emotion words (16,921 out of 28,969) would be ignored, i.e., if only the basic lexicon were used.
\section{Evaluation of the Constructed Lexicon}
\label{appendix:Eval}

Our out-of-sample testing, presented in the previous appendix, guarantees that the estimation of the algorithm is consistent with the basic lexicon, annotated by \cite{quan2010blog}. Nevertheless, one could be concerned that the basic lexicon might not be valid in our context. A more convincing evaluation would involve a comparison of our algorithm’s estimation to human annotations.

To do this, first, 1,200 newly mined emotion words were randomly selected as a testing set. Second, five research assistants were asked to independently annotate the emotional intensities for each of the 1,200 words and for each of the eight emotion categories. Third, we averaged the results provided by the annotators so that the results would not be biased due to any one of the annotators’ subjectivity. Then, for the eight emotion categories of the 1,200 words, we obtained 9,600 annotations by humans and 9,600 estimations by the algorithm. Finally, we conducted a Student’s t-test to compare the two results. We found that the difference between human annotation and algorithm estimation is insignificant ($p$ = 0.05, $N$ = 9,600).

\section{Parameter Selection for Topic Modeling}
\label{appendix:wechat}
\begin{figure}
  \begin{center}
      \includegraphics[width=0.6\textwidth]{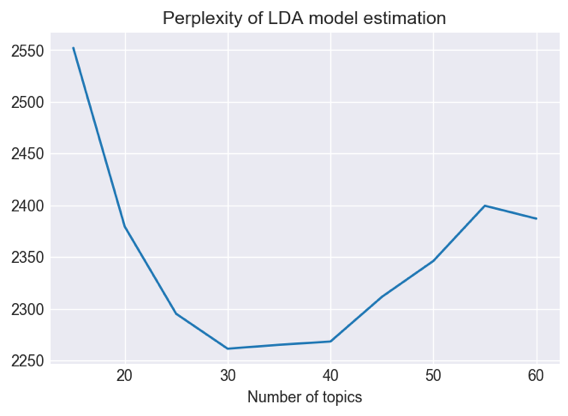}
  \end{center}
  
  \caption{Perplexity of Latent Dirichlet Allocation (LDA) model estimation across number of topics.}
  \label{fig:LDA}
\end{figure}

We used a Latent Dirichlet Allocation (LDA) topic model \citep{blei2003latent} and trained on article titles and content. Specifically, first, we used Chinese word segmentation repository by Python and cut titles and content into words. Second, we filtered out all emotion words (in our constructed lexicon) in these titles and content. By doing so, topic variables and emotion variables are constructed by different linguistic features. This reduces the possibility that the two sets of variables will confound each other. We also filtered out low-frequency words (i.e. words that appear in less than 0.1\% of the articles) to reduce noise in model estimation. Third, we trained an LDA model and specified the optimal number of topics that minimizes the model perplexity \citep{blei2003latent}. When the number of topics is 30, the perplexity is minimized (Figure \ref{fig:LDA}). Next, we used the trained LDA model to map the titles and content of each article to a 30-dimension vector as topic distribution, with each element as representing the probability that the article belongs to the corresponding topic category.

We present the results of parameter selection for topic modeling in Figure \ref{fig:LDA}. Perplexity is a commonly used metric to select the best parameter (i.e., number of topics) for an LDA model \citep{blei2003latent}. The Perplexity is defined as exp(-1 $\times$ log-likelihood per word). The optimal number of topics should minimize the perplexity of the model. Thus, we choose number of topics as 30.

\section{Effects of Positive Emotional Expressions} \label{appendix: emopos_results}
\subsection{Effects of Positive Emotional Expressions on Structural Properties of Cascades}
\label{appendix: emopos_main}
We controlled for the positive emotional expressions embedded in the articles when analyzing the effects of negative emotional expressions on cascade structural properties (i.e., depth, size, maximum breadth, and structural virality). Our results suggest that joy has a negative impact on cascade size ($\beta_{joy}^{size}=-0.003$, $p<0.1$), whereas surprise has a positive effect on cascade maximum breadth ($\beta_{surprise}^{breadth}=0.017$, $p<0.05$). We also find that articles with a higher degree of love lead to significantly higher cascade depth ($\beta_{love}^{depth}=0.006$, $p<0.01$), size ($\beta_{love}^{size}=0.004$, $p<0.01$), maximum breadth ($\beta_{love}^{breadth}=0.002$, $p<0.1$), and structural virality ($\beta_{love}^{sv}=0.004$, $p<0.01$).

\subsection{Effects of Positive Emotional Expressions on Demographics and Social Ties of Cascades}
\label{appendix: emopos_demo}
Similar to our analysis of the effects of negative emotional expressions on structural properties of cascades, we also controlled for the positive emotional expressions embedded in the articles when analyzing the effects of negative emotional expressions on demographics and social ties of cascades. Our results show that anticipation and love significantly and positively affect the average age of users who participate in the cascades ($\beta_{anticipation}^{Average Age}=0.079$, $p<0.001$, $\beta_{love}^{Average Age}=0.042$, $p<0.001$), whereas surprise has a negative effect on users' average age ($\beta_{surprise}^{Average Age}=-0.081$, $p<0.05$). Articles with a higher degree of joy and anticipation lead to a significantly lower proportion of females participating in the cascades ($\beta_{joy}^{Female Proportion}=-0.010$, $p<0.05$, $\beta_{anticipation}^{Female Proportion}=-0.013$, $p<0.1$), whereas articles with more expressions of love lead to a significantly higher proportion of female users engaged in the cascades ($\beta_{love}^{Female Proportion}=0.007$, $p<0.05$). Finally, surprise has a significant positive impact on the proportion of weak ties among users who participate in the cascades ($\beta_{surprise}^{Weak Tie Proportion}=0.073$, $p<0.05$). \clearpage

\section{Figures and Tables}
\subsection{Figures}
\label{appendix: Robust Figures}
%%%%%%%%%%%%%%%%%%%%%%%%%%%%%%%%%%%%%%%%%%%%%
%%%%%%%%%%%%%%%%%%%%%%%%%%%%%%%%%%%%%%%%%%%
\begin{figure}[H]
\begin{center}
    \includegraphics[width=0.85\columnwidth]{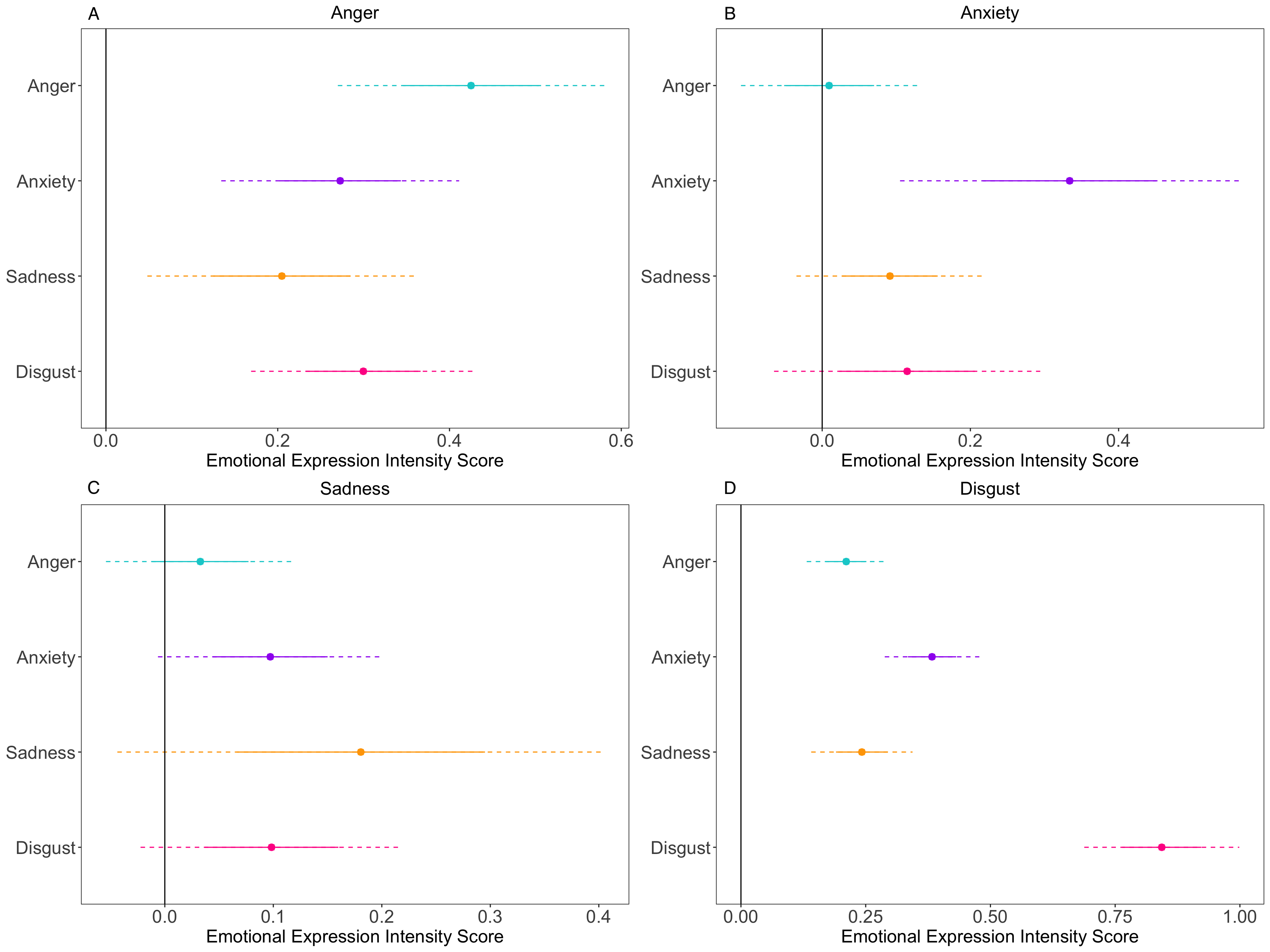}
\end{center}
\caption{Discrete emotions expressed in the comments of articles that contain extreme negative emotional expressions: (A) anger, (B) anxiety, (C) sadness, (D) disgust.}
\scriptsize Note: The $x$ axis indicates emotional expression intensity scores (after normalized to a standard normal distribution). The $y$ axis indicates the category of emotions in comments. The titles of the subplots indicate the type of the extreme emotion that the article contains. The articles that express an extreme emotion are those with only one of the emotion intensity scores higher than 1.96, whereas the others are lower than 1.96. The intervals indicated by dashed lines represent 95\% confidence intervals. The intervals indicated by solid lines represent mean values plus or minus 1 unit of their standard deviations. The overall figure indicates that the most-expressed emotion embedded in the comments is consistent with the mainly expressed emotion embedded in the articles.
\label{fig:CommentNew}
\end{figure}
% %%%%%%%%%%%%%%%%%%%%%%%%%%%%%%%%%%%%%%%%%%

\subsection{Tables}
\label{appendix: Robust Tables}
%%%%%%%%%%%%%%%%%%%%%%%%%%%%%%%%%%%%%%%%%%%%%%%%%%%%%%%%%%%%
%\begin{landscape}
\begin{table}[h]
\begin{center}
\begin{adjustbox}{width={0.8\textwidth},totalheight={\textheight},keepaspectratio}
% \begin{adjustbox}{max width=1.0\textwidth,center}
 \begin{threeparttable}
  \caption{Correlations among cascade dimensions\tnote{1}}
   \label{tab:cascadecor}
  \centering
  \begin{tabular}{l*{4}{D{.}{.}{-1}}}
  %\begin{tabularx}{\textwidth}{@{} l
  %*{5}{S}}
    \toprule
    \multicolumn{1}{c}{\textbf{ } }    & \multicolumn{1}{c}{\textbf{Depth}}     & \multicolumn{1}{c}{\textbf{Size}} & \multicolumn{1}{c}{\textbf{Maximum breadth}} & \multicolumn{1}{c}{\textbf{Structural virality}}\\
    \cmidrule(r){1-5}
    \textbf{Depth} &1.000	&		&		&	\\
    \textbf{Size} &0.720	&	1.000	&		&	\\
    \textbf{Maximum breadth} &0.588	&	0.974	&	1.000	&	\\   
    \textbf{Structural virality} &0.921	&	0.743	&	0.616&	1.000\\
    \bottomrule
  \end{tabular}
  %\begin{tablenotes}
  %\small
   % \item[] \textit{Note: (a) Given the right-skewed distributions of Size and Maximum Breadth, we did log-transformation on these two variables; (b) The topic distribution of the article is controlled; (c) Significance level: * $p < 0.05$, ** $p<0.01$, *** $p < 0.001$; (d) SV stands for structural virality.}
    %\end{tablenotes}
     \begin{tablenotes}
     \scriptsize
  \item[1] \textit{Cascades' size is positively correlated with maximum breadth and depth. Structural virality is positively correlated with cascades' depth, size, and maximum breadth.}
    \end{tablenotes}
\end{threeparttable}
\end{adjustbox}
\end{center}
\end{table}

%\end{landscape}

%%%%%%%%%%%%%%%%%%%%%%%%%%%%%%%%%%%%%%%%%%%%%%%%%%%%%%%%%%%%%%%%%
%%%%%%%%%%%%%%%%%%%%%%%%%%%%%%%%%%%%%%%%%%%%%%%%%%%%%%%%%%%%
%\begin{landscape}
\begin{table}[h]
\begin{center}
%\begin{adjustbox}{width={0.8\textwidth},totalheight={0.8\textheight},keepaspectratio}
 %\begin{adjustbox}{width={\textwidth},totalheight={\textheight},keepaspectratio}
 \begin{threeparttable}
  \caption{Correlations among Negative Emotions\tnote{1}}
   \label{tab:emocor}
  \centering
  \begin{tabular}{l*{4}{D{.}{.}{-1}}}
  %\begin{tabularx}{\textwidth}{@{} l
  %*{5}{S}}
    \toprule
    \multicolumn{1}{c}{\textbf{ } }    & \multicolumn{1}{c}{\textbf{Anger}}     & \multicolumn{1}{c}{\textbf{Anxiety}} & \multicolumn{1}{c}{\textbf{Sadness}} & \multicolumn{1}{c}{\textbf{Disgust}}\\
    %& \multicolumn{1}{c}{\textbf{Joy}} & \multicolumn{1}{c}{\textbf{Love}}& \multicolumn{1}{c}{\textbf{Surprise}}& \multicolumn{1}{c}{\textbf{Sadness}}\\
    \cmidrule(r){1-5}
    \textbf{Anger}&1.000\\
    \textbf{Anxiety}&0.190&	1.000\\
    \textbf{Sadness}&0.056&	0.178&1.000\\
    % \textbf{Anticipation}&-0.010&	0.109&	1.000\\   
    \textbf{Disgust}&0.435&	0.358&	0.058&	1.000\\
    % \textbf{Joy}&-0.065&	-0.001&	0.081&	-0.092&	1.000\\
    % \textbf{Love}&-0.138&	-0.153&	0.070&	-0.167&	0.294&	1.000\\
    % \textbf{Surprise}&0.139&	0.454&	-0.011&	0.247&	0.007&	-0.103&	1.000\\
    \bottomrule
  \end{tabular}
  \begin{tablenotes}
  \scriptsize
  \item[1] \textit{The absolute values of most correlations are below 0.435. The results indicate the independence of these four negative emotions.}
   \end{tablenotes}
\end{threeparttable}
%\end{adjustbox}
\end{center}
\end{table}
%\end{landscape}

%%%%%%%%%%%%%%%%%%%%%%%%%%%%%%%%%%%%%%%%%%%%%%%%%%%%%%%%%%%%%%%%%

%%%%%%%%%%%%%%%%%%%%%%%%%%%%%%%%%%%%%%%%%%%%%%%%%%%%%%%%%%%%%
%\begin{landscape}
\begin{table}[h]
 \begin{adjustbox}{width={1\textwidth},totalheight={1\textheight},keepaspectratio}
 \begin{threeparttable}
  \caption{Results: Excluding Articles with Videos and Few Words\tnote{1}}
   \label{tab:novideo}
  \centering
  \begin{tabular}{l*{5}{D{.}{.}{-1}}}
  %\begin{tabularx}{\textwidth}{@{} l
  %*{5}{S}}
    \toprule
    \multicolumn{1}{c}{\textbf{ } }    & \multicolumn{1}{c}{\textbf{Depth}}     & \multicolumn{1}{c}{\textbf{Size\tnote{2}}} & \multicolumn{1}{c}{\textbf{Breadth\tnote{2}}} & \multicolumn{1}{c}{\textbf{SV\tnote{5}}} \\
    \midrule
    \multicolumn{1}{l}{\textbf{Variable}}\\
    \cmidrule(r){1-5}
    Anger&-0.067*	&	-0.047*	&	-0.029*	&	-0.034	    \\
         & (0.039)\tnote{6}	&	(0.026)	&	(0.013)	&	(0.023)\\
    Anxiety&0.022***	&	0.016***	&	0.011***	&	0.014***\\
           &(0.006)	&	(0.004)	&	(0.003)	&	(0.003) \\
    Sadness&-0.03	&	-0.044*	&	-0.027*	&	-0.023\\
           &(0.054)	&	(0.013)	&	(0.010)	&	(0.031) \\
    Disgust&0.022**	&	0.009	&	0.002	&	0.011*\\
           &(0.010)	&	(0.006)	&	(0.006)	&	(0.006) \\
    % Joy&-0.004&     -0.003* &  -0.002 &  -0.002\\
    %   &(0.003) &   (0.002)&  (0.002)&  (0.002)\\
    % Love&0.006*** &   0.004*** & 0.002* & 0.004***\\
    %     &(0.002)  &  (0.001) & (0.001)&  (0.001) \\
    % Surprise&-0.009  &    0.009  & 0.017** &  -0.004 \\
    %         &(0.011)  &  (0.008)&  (0.008) & (0.007) \\
    % Anticipation&0.004 &     -0.003 &  -0.004 &  -0.001 \\
    %             & (0.005) &   (0.003) & (0.002)&  (0.003) \\
    \cmidrule(r){1-5}
    % \multicolumn{1}{l}{\textbf{Article-level Controls\tnote{3}}}\\
    % \cmidrule(r){1-6}
    % %Originality&0.093***&0.06***&0.051***&-0.052&0.056***\\
    % %Info. Uniqueness&-0.057***&-0.02***&0.021***&0.079&-0.032***\\
    % Article Length&0.008&0.034***&0.041***&-0.294***&-0.004\\
    % \# of Images&0.09***&0.042***&0.035***&-0.095*&0.05***\\
    % \# of Videos&0.163***&0.103***&0.096***&0.296***&0.081***\\
    % Post at Weekends&-0.004&0.006***&0.008***&0.064*&-0.001\\
    % \# of Comments&0.218***&0.114***&0.099***&-0.003&0.101***\\
    % Reward Function&0.03***&0.029***&0.028***&-0.106*&0.02***\\					
    % \cmidrule(r){1-6}
    % \multicolumn{1}{l}{\textbf{Publisher-level Controls}}\\
    % \cmidrule(r){1-6}
    % Ave. \# of Posts&-0.195***&0.035***&0.109***&-0.597***&-0.084***\\
    % \# of Followers&0.35***&0.575***&0.643***&1.484***&0.132***\\
    % Type: Individual&-0.009&0.016*&0.022**&0.252**&-0.005\\
    % Type: Business&-0.123***&-0.087***&-0.079***&0.348*&-0.089***\\
    % Type: Gov.&-0.066***&-0.022&-0.008&0.704***&-0.05***\\
    % Type: Media&-0.096*&-0.075&-0.055&0.449&-0.066**\\
    % \cmidrule(r){1-6}
    %\hline
    \textbf{Positive emotional expressions\tnote{3}}& \checkmark & \checkmark & \checkmark & \checkmark\cr
    \textbf{Article-level characteristics\tnote{4}}& \checkmark & \checkmark & \checkmark & \checkmark\cr
    \textbf{Publisher fixed effects}& \checkmark & \checkmark & \checkmark & \checkmark\cr
    \bottomrule
  \end{tabular}
  \begin{tablenotes}
  %\small
  \scriptsize
    \item[1] \textit{Significance level: * $p < 0.05$, ** $p<0.01$, *** $p < 0.001$}
    \item[2] \textit{Given the right-skewed distributions of size and maximum breadth, we did a log-transformation on these two variables.}
    \item[3] \textit{Positive emotional expressions, i.e., joy, love, surprise, and anticipation, are controlled.}
    \item[4] \textit{All of the article-level variables(i.e. article length, number of images and videos embedded in the article, whether the articles was posted during a weekend, number of comments and topic distribution of the article) are controlled.}
    \item[5] \textit{SV indicates structural virality.}
    \item[6] \textit{Standard error of estimation.}
    \end{tablenotes}
\end{threeparttable}
\end{adjustbox}
\end{table}
%\end{landscape}

% %%%%%%%%%%%%%%%%%%%%%%%%%%%%%%%%%%%%%%%%%%%%%%%%%%%%%%%%%%%%%
%%%%%%%%%%%%%%%%%%%%%%%%%%%%%%%%%%%%%%%%%%%%%%%%%%%%%%%%%%%%%
%\begin{landscape}
\begin{table}[h]
 \begin{adjustbox}{width={1\textwidth},totalheight={1\textheight},keepaspectratio}
 \begin{threeparttable}
  \caption{Results: Including Originality\tnote{1}}
   \label{tab:origin}
  \centering
  \begin{tabular}{l*{5}{D{.}{.}{-1}}}
  %\begin{tabularx}{\textwidth}{@{} l
  %*{5}{S}}
    \toprule
    \multicolumn{1}{c}{\textbf{ } }    & \multicolumn{1}{c}{\textbf{Depth}}     & \multicolumn{1}{c}{\textbf{Size\tnote{2}}} & \multicolumn{1}{c}{\textbf{Breadth\tnote{2}}} & \multicolumn{1}{c}{\textbf{SV\tnote{5}}} \\
    \midrule
    \multicolumn{1}{l}{\textbf{Variable}}\\
    \cmidrule(r){1-5}
    Anger&-0.069	&	-0.050*	&	-0.039*	&	-0.038	    \\
         &(0.042)\tnote{6}	&	(0.026)	&	(0.023)	&	(0.025)\\
    Anxiety&0.017***	&	0.013***	&	0.009***	&	0.009***\\
           &(0.006)	&	(0.004)	&	(0.003)	&	(0.003) \\
    Sadness&-0.047	&	-0.060*	&	-0.042	&	-0.037\\
           &(0.052)	&	(0.032)	&	(0.029)	&	(0.030) \\
    Disgust&0.021**	&	0.008	&	0.004	&	0.012**\\
           &(0.010)	&	(0.006)	&	(0.005)	&	(0.006)\\
    % Joy&-0.004&     -0.003* &  -0.002 &  -0.002\\
    %   &(0.003) &   (0.002)&  (0.002)&  (0.002)\\
    % Love&0.006*** &   0.004*** & 0.002* & 0.004***\\
    %     &(0.002)  &  (0.001) & (0.001)&  (0.001) \\
    % Surprise&-0.009  &    0.009  & 0.017** &  -0.004 \\
    %         &(0.011)  &  (0.008)&  (0.008) & (0.007) \\
    % Anticipation&0.004 &     -0.003 &  -0.004 &  -0.001 \\
    %             & (0.005) &   (0.003) & (0.002)&  (0.003) \\
    \cmidrule(r){1-5}
    % \multicolumn{1}{l}{\textbf{Article-level Controls\tnote{3}}}\\
    % \cmidrule(r){1-6}
    % %Originality&0.093***&0.06***&0.051***&-0.052&0.056***\\
    % %Info. Uniqueness&-0.057***&-0.02***&0.021***&0.079&-0.032***\\
    % Article Length&0.008&0.034***&0.041***&-0.294***&-0.004\\
    % \# of Images&0.09***&0.042***&0.035***&-0.095*&0.05***\\
    % \# of Videos&0.163***&0.103***&0.096***&0.296***&0.081***\\
    % Post at Weekends&-0.004&0.006***&0.008***&0.064*&-0.001\\
    % \# of Comments&0.218***&0.114***&0.099***&-0.003&0.101***\\
    % Reward Function&0.03***&0.029***&0.028***&-0.106*&0.02***\\					
    % \cmidrule(r){1-6}
    % \multicolumn{1}{l}{\textbf{Publisher-level Controls}}\\
    % \cmidrule(r){1-6}
    % Ave. \# of Posts&-0.195***&0.035***&0.109***&-0.597***&-0.084***\\
    % \# of Followers&0.35***&0.575***&0.643***&1.484***&0.132***\\
    % Type: Individual&-0.009&0.016*&0.022**&0.252**&-0.005\\
    % Type: Business&-0.123***&-0.087***&-0.079***&0.348*&-0.089***\\
    % Type: Gov.&-0.066***&-0.022&-0.008&0.704***&-0.05***\\
    % Type: Media&-0.096*&-0.075&-0.055&0.449&-0.066**\\
    % \cmidrule(r){1-6}
    %\hline
    \textbf{Positive emotional expressions\tnote{3}}& \checkmark & \checkmark & \checkmark & \checkmark\cr
    \textbf{Article-level characteristics\tnote{4}}& \checkmark & \checkmark & \checkmark & \checkmark\cr
    \textbf{Publisher fixed effects}& \checkmark & \checkmark & \checkmark & \checkmark\cr
    \bottomrule
  \end{tabular}
  \begin{tablenotes}
  %\small
  \scriptsize
    \item[1] \textit{Significance level: * $p < 0.05$, ** $p<0.01$, *** $p < 0.001$}
    \item[2] \textit{Given the right-skewed distributions of size and maximum breadth, we did a log-transformation on these two variables.}
    \item[3] \textit{Positive emotional expressions, i.e., joy, love, surprise, and anticipation, are controlled.}
    \item[4] \textit{All of the article-level variables(i.e. article length, number of images and videos embedded in the article, whether the articles was posted during a weekend, number of comments and topic distribution of the article) are controlled. Originality is also controlled.}
    \item[5] \textit{SV indicates structural virality.}
    \item[6] \textit{Standard error of estimation.}
    \end{tablenotes}
\end{threeparttable}
\end{adjustbox}
\end{table}
%\end{landscape}

% %%%%%%%%%%%%%%%%%%%%%%%%%%%%%%%%%%%%%%%%%%%%%%%%%%%%%%%%%%%%%
\clearpage
\end{document}